\documentclass[10pt,journal]{IEEEtran}
\IEEEoverridecommandlockouts
\usepackage{amsmath}
\usepackage{cite}
\usepackage{amsmath,amssymb,amsfonts}
\usepackage[ruled,linesnumbered,lined]{algorithm2e}
\usepackage{graphicx}
\usepackage{textcomp}
\usepackage{xcolor}
\usepackage{bm}
\usepackage{booktabs}
\usepackage{caption,subcaption}

\usepackage{booktabs}
\usepackage{tabularx}
\usepackage{array}
\usepackage{multirow}

\usepackage{stfloats}
\newcounter{MYtempeqncnt}

\usepackage{amsthm,amssymb,amsfonts}

\newtheorem{theorem}{Theorem}

\SetKwInput{Input}{input}
\SetKwInput{Output}{output}

\usepackage{array}
\newcolumntype{L}[1]{>{\raggedright\let\newline\\\arraybackslash\hspace{0pt}}m{#1}}
\newcolumntype{C}[1]{>{\centering\let\newline\\\arraybackslash\hspace{0pt}}m{#1}}
\newcolumntype{R}[1]{>{\raggedleft\let\newline\\\arraybackslash\hspace{0pt}}m{#1}}

\newtheorem{proposition}{Proposition}

\newtheorem{remark}{Remark}
\newtheorem{assumption}{Assumption}

\def\BibTeX{{\rm B\kern-.05em{\sc i\kern-.025em b}\kern-.08em
		T\kern-.1667em\lower.7ex\hbox{E}\kern-.125emX}}

\begin{document}
	\title{Two-Layer Microwave Linear Analog Computer (MiLAC)-Aided Multi-User MIMO Networks}
\author{Xiaohua Zhou, \textit{Graduate Student Member, IEEE}, Tianyu Fang, \textit{Graduate Student Member, IEEE}, Matteo Nerini, \textit{Senior Member, IEEE},
	  Yijie Mao, \textit{Senior Member, IEEE}, and Bruno Clerckx, \textit{Fellow, IEEE}
			\thanks{An earlier version of this paper has been accepted by the 2026 IEEE Global Communications Conference (GLOBECOM) \cite{two-layer-con}.
			\par This work has been supported by the National Nature Science Foundation of China under Grant 62571331. The work of Tianyu Fang was supported by the Research Council of Finland through 6G Flagship under Grant 346208 and through project DIRECTION under Grant 354901.
			\par X. Zhou and Y. Mao are with the School of Information Science and Technology, ShanghaiTech University, Shanghai 201210, China (e-mail:
			{zhouxh3, maoyj}@shanghaitech.edu.cn). 
			\par T. Fang is with Centre for Wireless Communications, University of Oulu, Finland (e-mail: tianyu.fang@oulu.fi).
			\par M. Nerini and 
			B. Clerckx are with the Department of Electrical and Electronic Engineering, Imperial College London, London SW7 2AZ, U.K. (e-mail: m.nerini20,  b.clerckx@imperial.ac.uk).
			}
		\\[-2.5 ex]
		\vspace{-0.8cm}
	}
	\maketitle
	\begin{abstract}
Microwave linear analog computer (MiLAC)-aided beamforming, which performs signal processing directly in the analog domain, has recently emerged as a promising alternative to fully digital and hybrid beamforming for multi-antenna systems. However, existing studies have shown that a single lossless and reciprocal MiLAC at the transmitter cannot, in general, attain the performance of fully digital beamforming in multi-user scenarios. To overcome this limitation, we propose a two-layer MiLAC-aided transmit beamforming architecture for downlink multi-user multiple-input multiple-output (MIMO) networks. Leveraging multiport microwave network theory, we first establish a general physics-compliant model of the proposed architecture, and then derive a simplified model under the assumptions of unilateral and perfectly matched amplifiers. We prove that, with lossless and reciprocal MiLACs, the two-layer MiLAC-aided transmitter can realize any fully digital transmit beamformer, while a single-layer MiLAC-aided receiver can attain the same achievable rate as a fully digital linear combiner. We then derive closed-form mappings from fully digital transmit beamformers and receive combiners to their corresponding MiLAC-aided analog implementations. Based on these equivalence results, we formulate a sum-rate maximization problem for the considered multi-user MIMO network and develop an optimization framework for jointly designing the power allocation and scattering matrices of the MiLAC-aided transceivers. Numerical results validate the theoretical analysis and show that the proposed two-layer MiLAC-aided design attains the same sum-rate performance as its fully digital counterpart.
	\end{abstract}
	
	\begin{IEEEkeywords}
		Microwave linear analog computer (MiLAC), multi-user multiple-input multiple-output (MIMO), sum-rate maximization.
	\end{IEEEkeywords}
	\section{Introduction}
	The rapid growth of wireless traffic, connected devices, and data-intensive applications is imposing increasingly strict requirements on future wireless networks in terms of data rate, spectral efficiency, coverage, and connectivity. To meet these demands, multiple-input multiple-output (MIMO) systems are evolving beyond the massive MIMO architectures deployed in fifth-generation (5G) networks toward ultra-massive, or \emph{gigantic}, MIMO systems comprising hundreds or even thousands of antennas. Such large-scale antenna arrays can provide substantial beamforming and spatial-multiplexing gains, making them particularly attractive for future wireless systems operating at higher carrier frequencies, where large array gains are essential to compensate for the more severe propagation losses \cite{giganticMIMO, OnTheRoad}.

The scalability of conventional MIMO transceiver architectures, however, becomes increasingly challenging as the number of antennas grows. In a \textit{fully digital beamforming} architecture, each antenna is connected to a dedicated radio-frequency (RF) chain comprising data converters, mixers, and other RF components. Although this architecture provides full beamforming flexibility by allowing the magnitude and phase of every beamforming coefficient to be independently controlled, its hardware cost and power consumption scale with the number of antennas. In addition, precoding and combining require digital matrix-vector operations on a per-symbol basis, resulting in a considerable computational burden for gigantic antenna arrays. \textit{Hybrid digital-analog beamforming} was introduced as a more scalable alternative, in which a low-dimensional digital beamformer is combined with an analog beamforming network, typically implemented using phase shifters \cite{ps-hbf,yu2016alternating}. By reducing the number of RF chains, hybrid beamforming substantially alleviates the hardware cost of fully digital architectures. Nevertheless, phase-shifter (PS)-based analog beamforming is subject to constant-modulus constraints and therefore offers less beamforming flexibility than fully digital beamforming. Moreover, since part of the signal processing remains in the digital domain, hybrid architectures still require digital precoding and combining, together with digital-to-analog and analog-to-digital conversion in the RF chains.

More recently, \textit{microwave linear analog computers (MiLACs)} have emerged as a fundamentally different approach to large-scale MIMO beamforming \cite{nerini2025milac1,nerini2025milac2,milac-mag}. A MiLAC is a reconfigurable multiport microwave network composed of tunable impedance elements to directly perform configurable linear transformations on RF signals as they propagate
through the network. At the transmitter, the data streams are supplied to the input ports of the MiLAC through RF chains, while its output ports are connected to the transmit antennas. At the receiver, a MiLAC can directly combine the signals collected by the antenna array before analog-to-digital conversion. Thus, precoding and combining are performed entirely in the analog domain rather than through symbol-by-symbol digital computations. This architecture offers several appealing features for gigantic MIMO. In particular, the number of RF chains can scale with the number of data streams rather than the number of antennas. Moreover, the RF chains only need to generate and detect constellation symbols, rather than precisely reproduce antenna-wise beamformed signals, allowing lower-resolution digital-to-analog converters (DACs) and analog-to-digital converters (ADCs). Computationally demanding beamforming operations can also be performed directly by the microwave network, eliminating the need for digital matrix-vector multiplication at every symbol interval~\cite{nerini2025milac2}.

The potential of MiLAC-aided communications has stimulated growing research interest. For point-to-point MIMO systems, a single layer of lossless and reciprocal MiLAC at both the transmitter and receiver was shown to remain capacity-achieving, i.e., to attain the same channel capacity as a fully digital transceiver \cite{nerini2025milac-capacity}. Several subsequent studies have further investigated the architectures and applications of MiLAC. In particular, the fully connected MiLAC architecture was shown to require a number of tunable impedance components that grows quadratically with the number of ports. This motivates the development of stem-connected MiLAC architectures \cite{nerini2025mimo}, inspired by stem-connected beyond diagonal reconfigurable intelligent
surface (BD-RIS) architectures \cite{qstem,wu2025band}. By exploiting a graph-theoretical representation of the microwave network, stem-connected MiLACs reduce the circuit-complexity from quadratic to linear scaling with the number of antennas while preserving the capacity-achieving property in point-to-point MIMO systems. 
Recent studies have addressed a range of practical considerations in MiLAC design, including channel estimation \cite{milac-ce, milac-ce-2026}, losses \cite{lossy-milac}, discrete elements \cite{stem-zhang,finite-milac}, mutual coupling \cite{physics-milac}, wideband effects \cite{wideband-milac,wideband-milac-journal}, energy efficiency \cite{milac-EE,ee-milac-all}, and impedance mismatch \cite{general-milac}.
MiLAC has also been studied in a broader range of applications, including multi-user communications \cite{multiuser-milac, limits-milac}, MIMO radar sensing and direction-of-arrival (DoA) estimation \cite{sensing-milac}, and joint active and passive beamforming \cite{passive-active-milac}. These advances underscore the potential of MiLAC as a versatile analog-computing platform for future large-scale wireless transceivers.

Despite this progress, an important limitation arises when one-layer MiLAC-aided beamforming is extended from point-to-point MIMO to multi-user communications. For a lossless and reciprocal MiLAC, the set of physically realizable beamforming matrices is smaller than that available to a fully digital architecture. Consequently, although a single-layer MiLAC is capacity-achieving in point-to-point MIMO systems, it cannot in general achieve the same beamforming flexibility as a fully digital transmitter in multi-user multiple-input single-output (MISO) systems \cite{multiuser-milac, limits-milac}. This limitation becomes particularly pronounced when the beamforming vectors required for different users are strongly correlated \cite{limits-milac}. While the performance gap may decrease in gigantic MIMO regimes as user channels become increasingly orthogonal, it remains severe in finite-dimensional multi-user systems.

One approach to overcoming this limitation is the \textit{hybrid digital-MiLAC} architecture \cite{limits-milac}. By introducing a low-dimensional digital beamformer before the MiLAC, this architecture achieves the same beamforming flexibility as a fully digital transmitter while using only as many RF chains as data streams. However, introducing such digital precoding sacrifices one of the key advantages of MiLAC-aided beamforming. The signals entering the RF chains are no longer simply the original constellation symbols but linear combinations generated by digital precoding. Consequently, higher-resolution DACs are
needed to accurately represent these signals, and per-symbol digital operations are reintroduced. 

Whether MiLAC architectures can achieve the full flexibility and performance of digital beamforming in general multi-user systems while remaining fully analog and using low-resolution data converters remains an open question. In this work, we propose a novel \textit{two-layer MiLAC} architecture comprising two lossless and reciprocal MiLAC networks interconnected through a power-allocation stage. The key finding is that this additional analog-processing layer overcomes the fundamental limitation of the one-layer architecture. For any fully digital precoding matrix, an equivalent two-layer MiLAC configuration can be constructed, thereby reproducing the same transmit beamforming operation without any digital precoder. It offers the beamforming flexibility of fully digital transmission while retaining key advantages of MiLAC, including a number of RF chains that scales with the number of transmitted streams, the potential use of low-resolution DACs, and fully analog symbol-level beamforming.

 The main contributions of this work are summarized as follows.
\begin{itemize}
	\item We propose a two-layer MiLAC architecture in which two reconfigurable microwave networks are interconnected through an active power-allocation stage. To the best of our knowledge, this is the first MiLAC architecture capable of realizing any fully digital precoding matrix in a multi-user MIMO system while performing beamforming entirely in the analog domain.
	\item We establish a physics-compliant model for the proposed two-layer architecture based on multiport microwave network theory, explicitly accounting for the interactions among the two MiLAC networks, the intermediate amplifiers, and their interconnections. Under the assumptions of unilateral and perfectly matched amplifiers, we further simplify this model to obtain a more tractable formulation.
	\item We analytically characterize the beamforming capabilities of the proposed two-layer MiLAC-aided transmitter and one-layer MiLAC-aided receiver. The different layer requirements arise because scaling the receive combiner does not affect the achievable rate, whereas scaling the transmit beamformer reduces the radiated power and may therefore lead to performance degradation under a fixed transmit-power budget. For any feasible fully digital precoding matrix, we show that there exists a two-layer MiLAC configuration that realizes exactly the same linear transformation. At the receiver, we show that a suitably scaled and conjugated fully digital minimum mean-square error (MMSE) combiner can be exactly implemented by a single-layer MiLAC without any loss in achievable rate. Closed-form constructions of the corresponding MiLAC scattering matrices are further derived. 
	\item We provide numerical results to validate the developed model and the theoretical analysis. The proposed two-layer architecture is compared with fully digital beamforming, one-layer MiLAC, hybrid digital-MiLAC, and PS-based hybrid beamforming under representative multi-user MIMO configurations. The results demonstrate that the proposed transceiver attains the same sum-rate performance as its fully digital counterpart while retaining the analog-processing advantages of MiLAC.
\end{itemize}
The remainder of this paper is organized as follows. Section~\ref{sec:proposed-architecture} introduces the physics-compliant and simplified model for the proposed two-layer MiLAC architecture. Section~\ref{Sec:system model} introduces the multi-user MIMO system model for a MiLAC-aided network. Section~\ref{sec:optimality} investigates the optimality of the proposed transceiver architecture. Section~\ref{sec:pro-alg} develops the corresponding problem formulation and algorithm design. Section~\ref{sec:sim} presents numerical results, and Section~\ref{sec:conclusion} concludes the paper.

\section{Proposed two-layer MiLAC architecture}\label{sec:proposed-architecture}
	In this section, we propose a two-layer MiLAC architecture comprising two MiLACs interconnected through a set of $D$ amplifiers. Specifically, we first develop a general physics-compliant model of the proposed architecture based on multiport microwave network theory. We then derive a simplified model under the assumptions of unilateral and perfectly matched amplifiers.
	
	\begin{figure}[!t]
		\centering
		\includegraphics[width=0.45\textwidth]{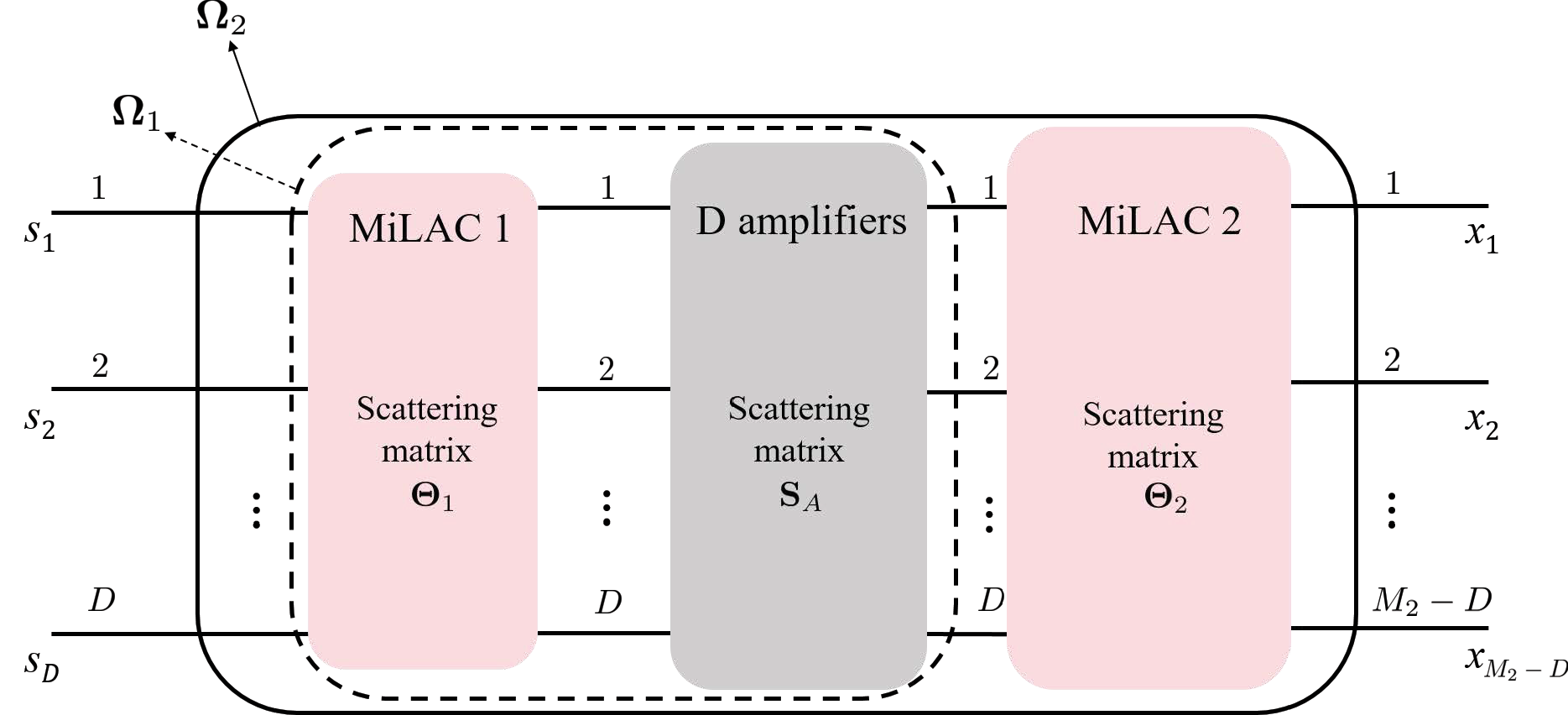}
		\caption{Block diagram of two-layer MiLAC-aided network. }
		\label{Fig:sys-model-S}
			\vspace{-0.5cm}
	\end{figure}

As depicted in Fig.~\ref{Fig:sys-model-S}, the proposed two-layer MiLAC-aided architecture receives $D$ input signals from $D$ RF chains and provides $L$ output signals to the transmit antennas. It comprises two cascaded MiLAC networks interconnected by $D$ parallel amplifiers. MiLAC 1 has $M_1=2D$ ports, comprising $D$ input ports connected to the RF chains and $D$ output ports connected to the amplifiers. MiLAC 2 has $M_2=D+L$ ports, comprising $D$ input ports connected to the amplifiers and $L$ output ports connected to the transmit antennas. The input vector $\mathbf{s}=[s_1,\ldots,s_D]^{\mathsf T}$ is first processed by MiLAC~1 in the analog domain. The resulting $D$ signals undergo amplitude adjustment and then are fed into MiLAC~2 for further analog processing. The resulting output vector is $\mathbf{x}=[x_1,\ldots,x_L]^{\mathsf T}$.


\subsection{MiLAC Modeling}\label{sec:MiLAC modeling}

According to microwave network theory, a multiport MiLAC can be characterized by a tunable admittance matrix whose entries depend on the network's adjustable admittance components. This representation applies to both the existing one-layer architecture and each MiLAC in the proposed two-layer architecture. Focusing on the latter, we denote the admittance matrix of the $m$-th MiLAC by $\mathbf{Y}_m\in\mathbb{C}^{M_m\times M_m}$, where $m\in\{1,2\}$ and $M_m$ denotes its number of ports.

Since the MiLACs considered in this work are fully connected, each port-$v$ of MiLAC $m$ is connected to ground through an admittance $Y_{m,v,v}\in\mathbb{C}$ and is interconnected with port-$i$ through an admittance $Y_{m,i,v}\in\mathbb{C}$ for all $i\neq v$. Let $\mathcal{V}_m = \{1,\ldots,M_m\}$ denote the port index set of MiLAC~$m$. As a function of these tunable admittance components, the $(i,v)$-th entry of $\mathbf{Y}_m$, for $i,v\in\mathcal{V}_m$, is given by
\begin{equation}\label{eq:admittance-matrix}
	[\mathbf{Y}_m]_{i,v}
	=
	\begin{cases}
		-Y_{m,i,v}, & i\neq v,\\[1mm]
		\displaystyle\sum_{j=1}^{M_m}Y_{m,j,v}, & i=v,
	\end{cases}
\end{equation}
which follows from Ohm's law \cite{nerini2025milac1}.

\begin{figure}[t]
	\centering
	\includegraphics[width=0.3\textwidth]{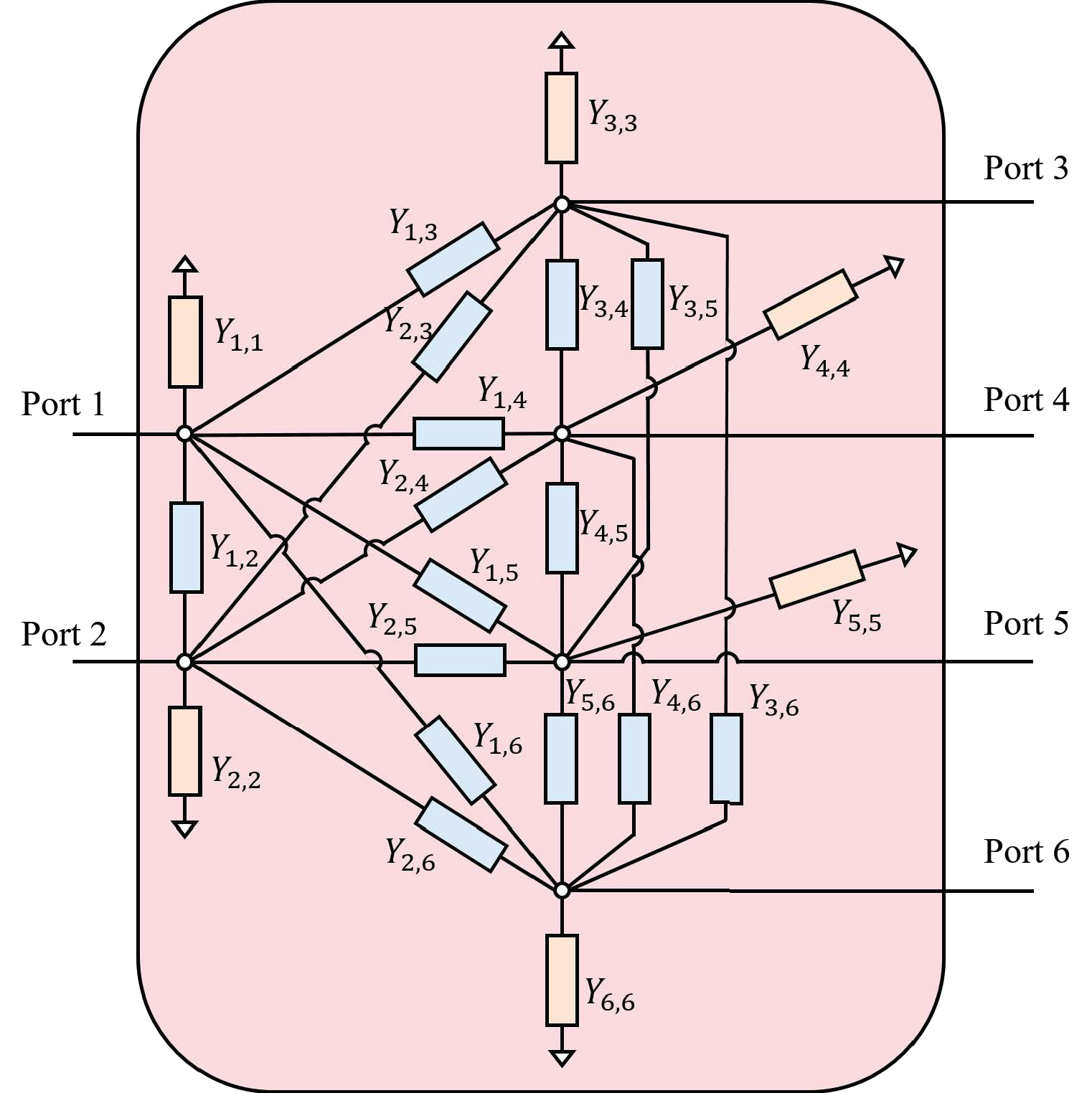}
	\caption{Example of lossless and reciprocal MiLAC with $ 6 $ ports. }
	\label{Fig:milac}
	\vspace{-0.6cm}
\end{figure}  
In the proposed two-layer MiLAC architecture, both MiLACs are assumed to be lossless and reciprocal. As an illustrative example, Fig.~\ref{Fig:milac} shows a fully connected lossless and reciprocal MiLAC with $2$ input ports and $4$ output ports. In the diagram, the blue admittances represent inter-port connections, while the orange admittances indicate ports connected to ground.
Under the lossless assumption, all tunable admittance components of both MiLACs are purely imaginary and can be written as $	Y_{m,i,v}=jB_{m,i,v}$, where $B_{m,i,v}\in\mathbb{R}$. Consequently, the admittance matrix of MiLAC $m$ can be expressed as $\mathbf{Y}_m=j\mathbf{B}_m$, where $\mathbf{B}_m=\mathbf{B}_m^{\mathsf T}$ and $
\mathbf{B}_m\in\mathbb{R}^{M_m\times M_m}$ is the corresponding susceptance matrix.

Let $\bm{\Theta}_m\in\mathbb{C}^{M_m\times M_m}$ denote the scattering matrix associated with MiLAC $m$. According to the relationship between the admittance and scattering matrices
\cite{nerini2025milac-capacity,nerini2025mimo,limits-milac}, we have
\begin{equation}
    \bm{\Theta}_m
    =
    \left(
    \mathbf I_{M_m}
    -j\frac{\mathbf B_m}{Y_0}
    \right)
    \left(
    \mathbf I_{M_m}
    +j\frac{\mathbf B_m}{Y_0}
    \right)^{-1}.
    \label{eq:cayley}
\end{equation}
Furthermore, the susceptance matrix $\mathbf B_m$ induces a
symmetric and unitary scattering matrix $\bm{\Theta}_m$. Accordingly, it satisfies
\begin{subequations}\label{eq:unified-MiLAC-constraints}
	\begin{align}
		\bm{\Theta}_m^{\mathsf H}\bm{\Theta}_m
		&=\mathbf{I}_{M_m},
		\label{eq:unitary-constraint}\\
		\bm{\Theta}_m
		&=\bm{\Theta}_m^{\mathsf T},
		\label{eq:symmetric-constraint}
	\end{align}
\end{subequations}
for all $m\in\{1,2\}$. Hence, MiLAC~1 and MiLAC~2 share the same mathematical representation and differ
only in their numbers of ports $M_m$.

\subsection{Amplifier Modeling}
Let $\mathcal{D}=\{1,\ldots,D\}$ denote the amplifier index set. The $d$-th amplifier is modeled as a linear two-port network \cite[Chapter 12.1]{pozar2021microwave} with scattering matrix
\begin{equation}
	\mathbf{S}_d
	=
	\begin{bmatrix}
		S_{d,11} & S_{d,12}\\
		S_{d,21} & S_{d,22}
	\end{bmatrix}
	\in\mathbb{C}^{2\times2},
	\qquad d\in\mathcal{D},
\end{equation}
where all scattering parameters are defined with respect to a common real reference impedance $Z_0$. Let $a_{d,1}$ and $a_{d,2}$ denote the incident waves at ports~1 and~2 of the $d$-th amplifier, respectively, while $b_{d,1}$ and $b_{d,2}$ denote the corresponding outgoing waves. Their scattering relation is given by
\begin{equation}
	\begin{bmatrix}
		b_{d,1}\\
		b_{d,2}
	\end{bmatrix}
	=
	\mathbf{S}_d
	\begin{bmatrix}
		a_{d,1}\\
		a_{d,2}
	\end{bmatrix}.
\end{equation}
Here, $S_{d,11}$ and $S_{d,22}$ are the input and output reflection coefficients, respectively, while $S_{d,21}$ and $S_{d,12}$ denote the forward and reverse transmission coefficients, respectively.


To model the $D$ parallel amplifiers as a single network, we collect the incident and outgoing waves at each port across all branches as
\begin{equation}
\begin{split}
		\mathbf{a}_m
	=&
	\begin{bmatrix}
		a_{1,m}&\cdots&a_{D,m}
	\end{bmatrix}^{\mathsf T},\\
	\mathbf{b}_m
	=&
	\begin{bmatrix}
		b_{1,m}&\cdots&b_{D,m}
	\end{bmatrix}^{\mathsf T},
	m\in\{1,2\}.
\end{split}
\end{equation}
The amplifier bank is therefore characterized by
\begin{equation}
	\begin{bmatrix}
		\mathbf{b}_1\\
		\mathbf{b}_2
	\end{bmatrix}
	=
	\underbrace{
		\begin{bmatrix}
			\mathbf{S}_{A,11} & \mathbf{S}_{A,12}\\
			\mathbf{S}_{A,21} & \mathbf{S}_{A,22}
	\end{bmatrix}}_{\mathbf{S}_A}
	\begin{bmatrix}
		\mathbf{a}_1\\
		\mathbf{a}_2
	\end{bmatrix},
\end{equation}
where $\mathbf{S}_A\in\mathbb{C}^{2D\times2D}$. Since the amplifier branches operate independently, the four submatrices are diagonal and given by
\begin{equation}
	\mathbf{S}_{A,mn}
	=
	\operatorname{diag}
	\left(S_{1,mn},\ldots,S_{D,mn}\right),
	\quad m,n\in\{1,2\}.
\end{equation}

\subsection{End-to-End Modeling of the Two-Layer MiLAC Network}
The proposed two-layer MiLAC architecture can be viewed as a cascade of three multiport networks: MiLAC~1, $D$ amplifiers, and MiLAC~2. According to the MiLAC modeling in Section \ref{sec:MiLAC modeling}, we partition the scattering matrices of MiLAC $m$ as 
\begin{equation}\label{eq:Theta-partition}
	\bm\Theta_m=
	\begin{bmatrix}
		\bm\Theta_{m,11} & \bm\Theta_{m,21}^\mathsf{T}\\
		\bm\Theta_{m,21} & \bm\Theta_{m,22}
	\end{bmatrix},
\end{equation}
where $\bm\Theta_{m,11}\in\mathbb{C}^{D\times D}$, $\bm\Theta_{m,21}\in\mathbb{C}^{(M_m-D)\times D}$, and $\bm\Theta_{m,22}\in\mathbb{C}^{(M_m-D)\times (M_m-D)}$. To characterize the cascade of these multiport networks while accounting for the interactions among their interconnected ports, we first introduce the following proposition.
\begin{figure}[!t]
	\centering
	\includegraphics[width=0.4\textwidth]{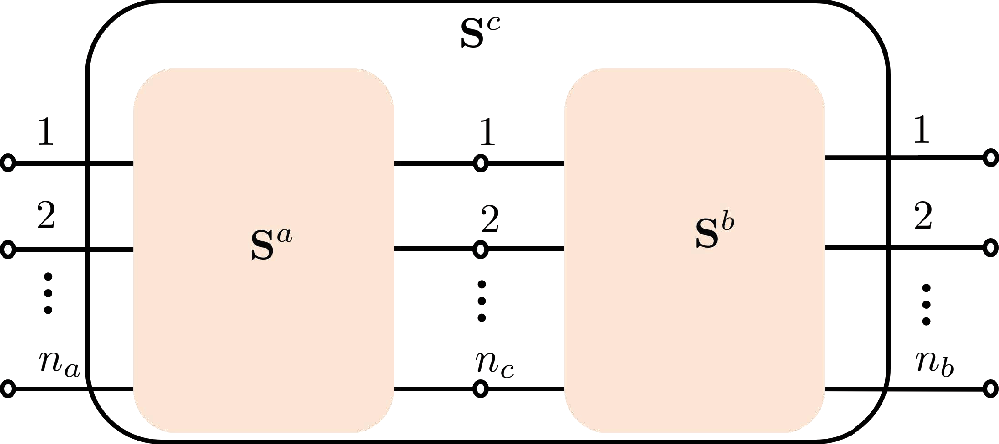}
	\caption{Cascade of two multiport networks. }
	\label{Fig:cascade}
		\vspace{-0.4cm}
\end{figure} 

\begin{figure*}[t!]
	\normalsize
	\setcounter{MYtempeqncnt}{\value{equation}}
	\setcounter{equation}{17}
	
	\vspace*{2pt}
	
	\begin{equation}\label{eq:beamforming-whole}
		\begin{split}
			\mathbf{x}
			=&\frac{1}{2}
			\bm{\Omega}_{2,21}\mathbf{s}\\
			=&\frac{1}{2}
			\bm\Theta_{2,21}
			\left(
			\mathbf{I}-
			\left(
			\mathbf{S}_{A,22}
			+
			\mathbf{S}_{A,21}
			\left(
			\mathbf{I}
			-
			\bm\Theta_{1,22}\mathbf{S}_{A,11}
			\right)^{-1}
			\bm\Theta_{1,22}\mathbf{S}_{A,12}
			\right)
			\bm\Theta_{2,11}
			\right)^{-1}
			\mathbf{S}_{A,21}
			\left(
			\mathbf{I}
			-
			\bm\Theta_{1,22}\mathbf{S}_{A,11}
			\right)^{-1}
			\bm\Theta_{1,21}\mathbf{s}.
		\end{split}
	\end{equation}
	\hrulefill
	\setcounter{equation}{\value{MYtempeqncnt}}
\end{figure*}	
\begin{proposition}\label{pro:cascade-net}
	Consider two multiport networks connected in cascade as illustrated in Fig. \ref{Fig:cascade}. Let the first network be an $N_a$-port network, where $N_a=n_a+n_c$, with scattering matrix
	\begin{equation}
		\mathbf{S}^a=
		\begin{bmatrix}
			\mathbf{S}^a_{11} & \mathbf{S}^a_{12}\\
			\mathbf{S}^a_{21} & \mathbf{S}^a_{22}
		\end{bmatrix},
	\end{equation}
	where $\mathbf{S}^a_{11}\in\mathbb{C}^{n_a\times n_a}$ and $\mathbf{S}^a_{22}\in\mathbb{C}^{n_c\times n_c}$. Let the second network be an $N_b$-port network, where $N_b=n_c+n_b$, with scattering matrix
	\begin{equation}
		\mathbf{S}^b=
		\begin{bmatrix}
			\mathbf{S}^b_{11} & \mathbf{S}^b_{12}\\
			\mathbf{S}^b_{21} & \mathbf{S}^b_{22}
		\end{bmatrix},
	\end{equation}
		where $\mathbf{S}^b_{11}\in\mathbb{C}^{n_c\times n_c}$ and $\mathbf{S}^b_{22}\in\mathbb{C}^{n_b\times n_b}$. Suppose that the last $n_c$ ports of the first network are connected one-to-one to the first $n_c$ ports of the second network. Assuming that the matrix inverses appearing below exist, the resulting interconnected network is an $N_c$-port network, where $N_c=n_a+n_b$, with scattering matrix
	\begin{equation}
		\mathbf{S}^c=
		\begin{bmatrix}
			\mathbf{S}^c_{11} & \mathbf{S}^c_{12}\\
			\mathbf{S}^c_{21} & \mathbf{S}^c_{22}
		\end{bmatrix},
	\end{equation}
	whose submatrices are given by
	\begin{subequations}
		\begin{align}
			\mathbf{S}^c_{11}
			=&
			\mathbf{S}^a_{11}
			+
			\mathbf{S}^a_{12}
			\left(\mathbf{I}-\mathbf{S}^b_{11}\mathbf{S}^a_{22}\right)^{-1}
			\mathbf{S}^b_{11}\mathbf{S}^a_{21},\\
			\mathbf{S}^c_{12}
			=&
			\mathbf{S}^a_{12}
			\left(\mathbf{I}-\mathbf{S}^b_{11}\mathbf{S}^a_{22}\right)^{-1}
			\mathbf{S}^b_{12},\\
				\mathbf{S}^c_{21}
			=&
			\mathbf{S}^b_{21}
			\left(\mathbf{I}-\mathbf{S}^a_{22}\mathbf{S}^b_{11}\right)^{-1}
			\mathbf{S}^a_{21},\\
			\mathbf{S}^c_{22}
			=&
			\mathbf{S}^b_{22}
			+
			\mathbf{S}^b_{21}
			\left(\mathbf{I}-\mathbf{S}^a_{22}\mathbf{S}^b_{11}\right)^{-1}
			\mathbf{S}^a_{22}\mathbf{S}^b_{12}.
		\end{align}
	\end{subequations}
\end{proposition}

\textit{Proof:}  
The detailed proof follows the procedure in \cite[Appendix]{matteo-SIM}.
\hfill$\square$

Applying Proposition~\ref{pro:cascade-net} to the cascade of MiLAC~1 and $D$ amplifiers, we obtain the scattering matrix 
\begin{equation}
	\bm{\Omega}_1 =
	\begin{bmatrix}
		\bm{\Omega}_{1,11} & \bm{\Omega}_{1,12}\\
		\bm{\Omega}_{1,21} & \bm{\Omega}_{1,22}
	\end{bmatrix},
\end{equation}
with
\begin{subequations}
	\begin{align}
		\bm{\Omega}_{1,11}
		=&
		\bm\Theta_{1,11}
		+
		\bm\Theta_{1,21}^\mathsf{T}
		\left(\mathbf{I}-\mathbf{S}_{A,11}\bm\Theta_{1,22}\right)^{-1}
		\mathbf{S}_{A,11}\bm\Theta_{1,21},\\
		\bm{\Omega}_{1,12}
		=&
		\bm\Theta_{1,21}^\mathsf{T}
		\left(\mathbf{I}-\mathbf{S}_{A,11}\bm\Theta_{1,22}\right)^{-1}
		\mathbf{S}_{A,12},
\\
		\bm{\Omega}_{1,21}
		=&
		\mathbf{S}_{A,21}
		\left(\mathbf{I}-\bm\Theta_{1,22}\mathbf{S}_{A,11}\right)^{-1}
		\bm\Theta_{1,21},
\\
		\bm{\Omega}_{1,22}
		=&
		\mathbf{S}_{A,22}
		+
		\mathbf{S}_{A,21}
		\left(\mathbf{I}-\bm\Theta_{1,22}\mathbf{S}_{A,11}\right)^{-1}
		\bm\Theta_{1,22}\mathbf{S}_{A,12}.
	\end{align}
\end{subequations}
Cascading the resulting network with MiLAC~2 and
applying Proposition~\ref{pro:cascade-net} again yields
\begin{equation}
	\bm{\Omega}_2 =
	\begin{bmatrix}
		\bm{\Omega}_{2,11} & \bm{\Omega}_{2,12}\\
		\bm{\Omega}_{2,21} & \bm{\Omega}_{2,22}
	\end{bmatrix},
\end{equation}
with
\begin{subequations}
	\begin{align}
		\bm{\Omega}_{2,11}
		=&
		\bm{\Omega}_{1,11}
		+
		\bm{\Omega}_{1,12}
		\left(\mathbf{I}-\bm \Theta_{2,11}\bm{\Omega}_{1,22}\right)^{-1}
		\bm \Theta_{2,11}\bm{\Omega}_{1,21},
\\
		\bm{\Omega}_{2,12}
		=&
		\bm{\Omega}_{1,12}
		\left(\mathbf{I}-\bm \Theta_{2,11}\bm{\Omega}_{1,22}\right)^{-1}
		\bm\Theta_{2,21}^\mathsf{T},
\\
		\bm{\Omega}_{2,21}
		=&
		\bm \Theta_{2,21}
		\left(\mathbf{I}-\bm{\Omega}_{1,22}\bm \Theta_{2,11}\right)^{-1}
		\bm{\Omega}_{1,21},
\\
		\bm{\Omega}_{2,22}
		=&
		\bm \Theta_{2,22}
		+
		\bm \Theta_{2,21}
		\left(\mathbf{I}-\bm{\Omega}_{1,22}\bm \Theta_{2,11}\right)^{-1}
		\bm{\Omega}_{1,22}\bm\Theta_{2,21}^\mathsf{T}.
	\end{align}		
\end{subequations}

Hence, the complete two-layer architecture can be represented as a single composite multiport network in Fig. \ref{Fig:sys-model-S}. The inner dashed boundary encloses MiLAC 1 and the amplifier bank, with composite scattering matrix $\bm{\Omega}_1$. The outer solid boundary encloses the complete two-layer network, with scattering matrix $\bm{\Omega}_2$. The corresponding end-to-end relationship between the input $\mathbf{s}$ and output $\mathbf{x}$ is given by \eqref{eq:beamforming-whole}.

\begin{figure*}[t!]
	\centering
	\includegraphics[width=0.6\textwidth]{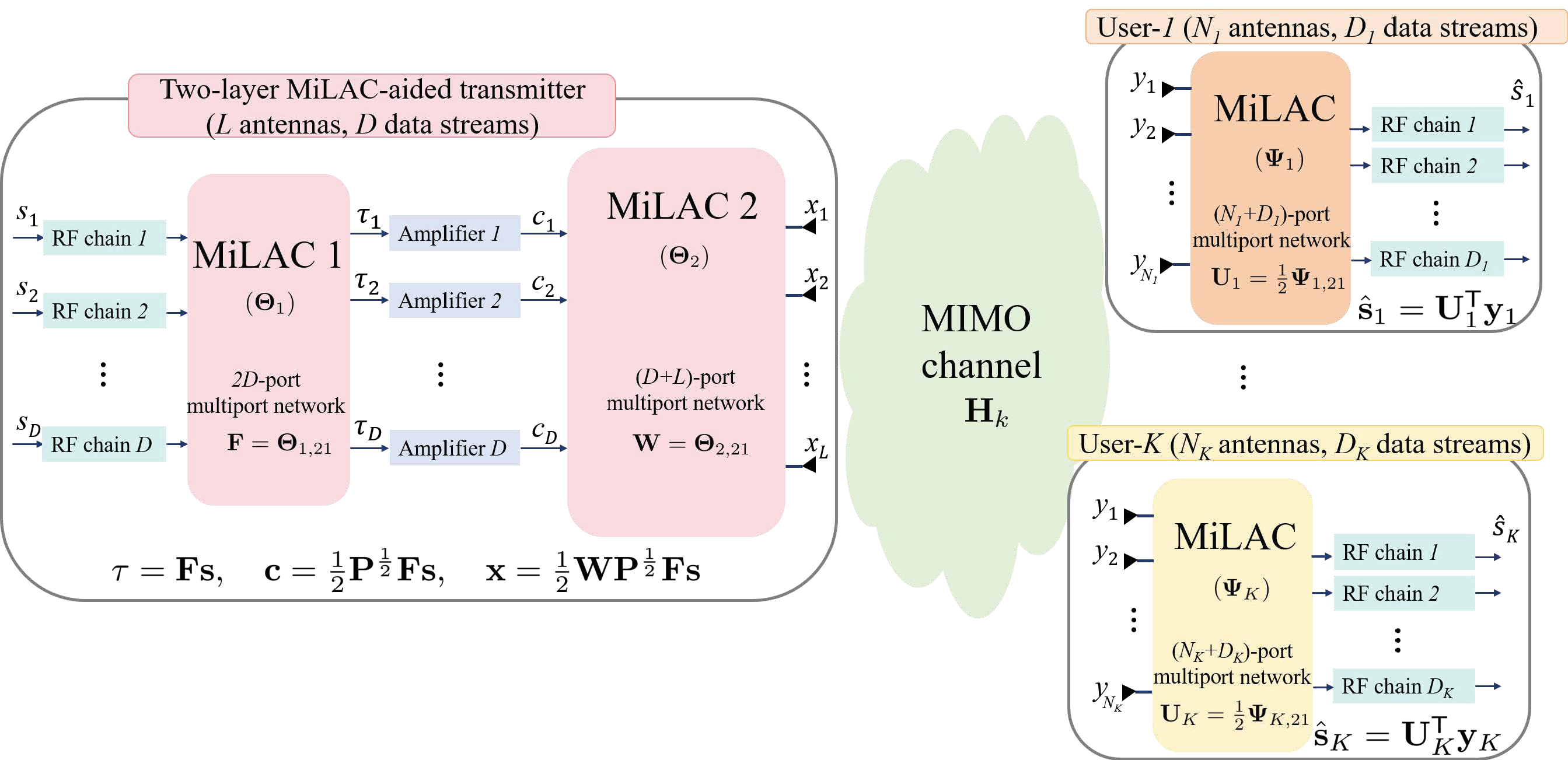}
	\caption{Proposed MiLAC-aided downlink multi-user MIMO architecture with a two-layer MiLAC at the transmitter and one-layer MiLAC at the receiver. }
	\label{Fig:sys-transceiver}
		\vspace{-0.5cm}
\end{figure*} 

\stepcounter{equation}
To obtain a tractable representation of the proposed two-layer MiLAC architecture, we adopt the following two assumptions on the intermediate amplifiers.
\begin{assumption}[Unilateral amplifiers]\label{assu:uni}
	Each amplifier is assumed to be unilateral, i.e., there is no reverse transmission from its output port to its input port.
\end{assumption}
Under Assumption \ref{assu:uni}, we have $S_{d,12}=0, \forall d\in\mathcal{D}$ or, equivalently, $\mathbf{S}_{A, 12}=\mathbf{0}_{D\times D}$. Accordingly, \eqref{eq:beamforming-whole} simplifies to
\begin{equation}\label{eq:re-assum1}
		\mathbf{x}
		=\frac{1}{2}
		\bm \Theta_{2,21} \bm \Lambda
		\bm\Theta_{1,21} \mathbf{s},
\end{equation}
where $\bm \Lambda =\left(\mathbf{I}-\mathbf{S}_{A,22}\bm\Theta_{2,11}\right)^{-1}
\mathbf{S}_{A,21}
\left(\mathbf{I}-\bm\Theta_{1,22}\mathbf{S}_{A,11}\right)^{-1}$.	
The unilateral assumption provides a useful approximation for many transistor devices~\cite[Sec.~12.1]{pozar2021microwave}.
\begin{assumption}[Perfectly matched amplifiers]\label{assu:perf}
	The input and output ports of each amplifier are assumed to be perfectly matched to the reference impedance $Z_0$.
\end{assumption}
 Under both Assumption \ref{assu:uni} and \ref{assu:perf}, we have the corresponding input and output reflection coefficients of each amplifier vanish, i.e., $ S_{d,11}=S_{d,22}=0, \forall d\in\mathcal{D}$, which gives $ \mathbf{S}_{A, 11} = \mathbf{S}_{A, 22} = \mathbf{0}_{D\times D}$. Therefore, the scattering matrix $\mathbf{S}_A$ for the $D$ amplifiers becomes
\begin{equation}
	\mathbf{S}_A=\begin{bmatrix}
		\mathbf{0}&\mathbf{0}\\
		\mathbf{S}_{A,21}&\mathbf{0}
	\end{bmatrix}.
\end{equation}
Substituting these assumptions into the input–output relationship of the transmitter in \eqref{eq:beamforming-whole} yields
	\begin{equation}\label{eq:simplified}
	\begin{split}
		\mathbf{x}= \frac{1}{2}
		\bm \Theta_{2,21}
		\mathbf{S}_{A,21}
		\bm\Theta_{1,21} \mathbf{s}.
	\end{split}
\end{equation}	
In the remainder of this paper, we adopt the simplified
model in \eqref{eq:simplified} for transceiver analysis and design.

\begin{remark}
 Figs. \ref{Fig:sys-model-S} and \ref{Fig:sys-transceiver} show an equivalent representation of the proposed architecture and do not prescribe the physical placement of conventional power amplifiers (PAs). In the simplified model in \eqref{eq:simplified}, the intermediate stage implements the nonuniform amplitude weights required to distribute power among the input ports of MiLAC~2. The diagonal gain matrix can be factored as $\mathbf{S}_{A,21}=\alpha \mathbf{A}$, where $\alpha>0$ is a common amplitude gain and $\mathbf{A}$ is a diagonal attenuation matrix. Thus, \eqref{eq:simplified} can be equivalently expressed as 
 \begin{equation*}
 	\begin{split}
 		\mathbf{x} =& \frac{1}{2}
 		\bm \Theta_{2,21}
 		\left(\alpha \mathbf{A}\right)
 		\bm\Theta_{1,21} \mathbf{s}\\
 		=& \frac{1}{2}
 		\left(\alpha \bm \Theta_{2,21}\right) 
 	\mathbf{A}
 		\bm\Theta_{1,21} \mathbf{s}\\
 		=& \frac{1}{2}
 		\bm \Theta_{2,21}
 		\mathbf{A}
 		 \left(\alpha \bm\Theta_{1,21}\right) \mathbf{s}.
 	\end{split}
 \end{equation*}
 Since the scalar gain $\alpha$ commutes with both MiLAC transmission matrices, conventional PAs providing this common gain may be placed before MiLAC~1, between the two MiLAC layers, or after MiLAC~2 without changing the input-output relationship in \eqref{eq:simplified}. In contrast, $\mathbf{A}$ generally does not commute with the MiLAC transmission matrices and is therefore retained between the two layers to preserve the required nonuniform amplitude weighting. The physical placement of the PAs remains subject to hardware constraints, including power handling, impedance matching, and PA linearity.
\end{remark}

	\section{MiLAC-Aided Multi-User MIMO System Model}\label{Sec:system model}
	
	Building on the simplified transmitter input–output relationship in \eqref{eq:simplified}, we now develop a system-level model for a narrowband downlink multi-user MIMO system employing the proposed MiLAC architecture. As illustrated in Fig.~\ref{Fig:sys-transceiver}, the base station employs a two-layer MiLAC transmitter, while each user employs a single-layer MiLAC receiver. This asymmetric configuration follows from the different requirements of transmit precoding and receive combining: a single-layer MiLAC can realize an arbitrary receive combiner up to a scaling factor that does not affect the achievable rate, whereas two MiLAC layers are generally required to realize an arbitrary fully digital precoder without a transmit-power loss.

	The transmitter is equipped with $L$ antennas and simultaneously serves $K$ users by transmitting $D$ data streams in parallel, while user $k$ is equipped with $N_k$ receive antennas and requires $D_k$ independent data streams. Here, we assume $D \leq \min \{L, \sum_{k =1}^K N_k\}$ and $D_k \leq \min \{L, N_k\}$. Let $\mathbf{s}_k \in \mathbb{C}^{D_k \times 1}$ denote the data symbol vector intended for user $k$, where $
	\mathbf{s}_k \sim \mathcal{CN}\left(\mathbf{0},\mathbf{I}_{D_k}\right)$. The aggregate data symbol vector for all users is given by 
	$\mathbf{s}
	=
	\begin{bmatrix}
		\mathbf{s}_1^{T},
		\mathbf{s}_2^{T},
		\ldots,
		\mathbf{s}_K^{T}
	\end{bmatrix}^{T}
	\in \mathbb{C}^{D \times 1}$,
	where $D=\sum_{k=1}^{K}D_k$ denotes the total number of transmitted data streams. Assuming that the data streams are mutually independent, we have $
	\mathbf{s}\sim\mathcal{CN}\left(\mathbf{0},\mathbf{I}_D\right)$.

	At the transmitter, beamforming is implemented using the proposed two-layer MiLAC-aided architecture with $M_1=2D$, $M_2=L+D$. First, the $D$ data streams intended for the users are processed in parallel by $D$ RF chains, which convert the digital symbols into analog RF signals before they are fed into MiLAC~1. As discussed earlier, the MiLAC architecture allows the RF chains to directly process the original constellation symbols, thereby enabling the use of low-resolution DACs. The outputs of the RF chains are then fed into the input ports of MiLAC~1, which performs the first-stage analog precoding. Subsequently, the $D$ output signals of MiLAC~1 are passed through $D$ parallel amplifiers located between MiLAC~1 and MiLAC~2, which implement power allocation. Finally, MiLAC~2 performs the second-stage analog precoding, thereby mapping the $D$ input signals to the $L$ transmit antennas.

\subsection{Transmitter Model}

With the proposed two-layer MiLAC architecture, the output signal of MiLAC 1 is given by
 	\begin{equation}
 		\bm{\tau} = \mathbf{F}\mathbf{s},
 	\end{equation}
 	where $\mathbf{F} \in \mathbb{C}^{D \times D}$ denotes the first-stage analog beamforming matrix. Based on the relationship between the scattering matrix and the MiLAC-aided beamforming matrix~\cite{nerini2025milac-capacity,nerini2025mimo,limits-milac}, we have
 	\begin{equation}\label{eq:theta1}
 		\mathbf{F}
 		=
 		\bm{\Theta}_{1,21}.
 	\end{equation}
 Note that $\mathbf F$ is defined directly in terms of the transmission block of the scattering matrix. The normalization factor associated with the overall MiLAC input-output relation is accounted for in the subsequent definition of the amplifier power-allocation coefficients.

 	Let $\mathbf{P}
 	=
 	\operatorname{diag}(p_1,p_2,\ldots,p_D)
 	\in \mathbb{R}^{D \times D}$
 	denote the power allocation matrix associated with the $D$ amplifiers between MiLAC 1 and MiLAC 2, where $p_d \geq 0$ represents the power allocated to the $d$-th intermediate signal stream. After amplitude adjustment and power allocation, the input signal to MiLAC 2 is given by
 	\begin{equation}
 		\mathbf{c}
 		=
 		\frac{1}{2}\mathbf{P}^{\frac{1}{2}}\bm{\tau}
 		=
 		\frac{1}{2}\mathbf{P}^{\frac{1}{2}}\mathbf{F}\mathbf{s},
 	\end{equation}
 	where $	\mathbf{P}^{\frac{1}{2}}
 	=
 	\operatorname{diag}\left(
 	\sqrt{p_1},\sqrt{p_2},\ldots,\sqrt{p_D}
 	\right)
 	\in \mathbb{R}^{D \times D}$ 
 	is the principal square root of the power allocation matrix and $p_d=|S_{d,21}|^2\geq0$\footnote{We assume that the power amplifiers do not introduce any phase shift to the signal, i.e., $\arg(S_{d,21})=0$. Even if phase shifts are introduced, we assume that they are known and can be compensated by reconfiguring one of the two MiLACs accordingly.} denotes the power-allocation coefficient associated with the $d$-th amplifier. The power allocation coefficients satisfy $\sum_{d=1}^{D} p_d \leq P_t$, where $P_t$ denotes the transmit power budget. Finally, after second-stage analog beamforming by MiLAC 2, the transmit signal vector at the transmitter is
 	\begin{equation}\label{eq:x-s-beamforming}
 		\mathbf{x}
 		=
 		\mathbf{W}\mathbf{c}
 		=
 		\frac{1}{2}\mathbf{W}\mathbf{P}^{\frac{1}{2}}\mathbf{F}\mathbf{s},
 	\end{equation}
 	where $\mathbf{W} \in \mathbb{C}^{L \times D}$ denotes the second-stage analog beamforming matrix. Similarly to \eqref{eq:theta1}, the analog beamforming matrix $\mathbf{W}$ implemented by MiLAC~2 is given by
 	\begin{equation}\label{eq:theta2}
 		\mathbf{W}
 		=
 		\bm{\Theta}_{2,21}.
 	\end{equation}
 
Comparing \eqref{eq:simplified} and \eqref{eq:x-s-beamforming}, the forward scattering matrix of the $D$ amplifiers is related to the power-allocation matrix as
\begin{equation}\label{eq:amplifier-gain}
   \mathbf{P}^{\frac{1}{2}}=\mathbf{S}_{A,21}.
\end{equation}

\subsection{Receiver Model}
	  The signal $\mathbf{y}_k \in \mathbb{C}^{N_k \times 1}$ received at user $k$ is given by
	\begin{equation}
		\begin{split}
			\mathbf{y}_k
			=&\mathbf{H}_k\mathbf{x}+\mathbf{n}_k\\
			=&\frac{1}{2}\mathbf{H}_k\mathbf{W}\mathbf{P}^{\frac{1}{2}} \mathbf{F}\mathbf{s}+\mathbf{n}_k\\
			=&\frac{1}{2}\mathbf{H}_k\mathbf{W}\mathbf{P}^{\frac{1}{2}} \mathbf{F}_k\mathbf{s}_k +\frac{1}{2} \sum_{i \neq k}\mathbf{H}_k\mathbf{W}\mathbf{P}^{\frac{1}{2}} \mathbf{F}_i\mathbf{s}_i+\mathbf{n}_k,
		\end{split}
	\end{equation}
	where $\mathbf{H}_k\in\mathbb{C}^{N_k\times L}$ denotes the channel matrix from the transmitter to user~$k$ and $\mathbf{n}_k\sim\mathcal{CN}
(\mathbf{0},\sigma_k^2\mathbf{I}_{N_k})$ denotes the additive noise. Moreover, $\mathbf{F}_k\in\mathbb{C}^{D\times D_k}$ is the submatrix of $\mathbf{F}$ associated with user~$k$, such that $\mathbf{F}=[\mathbf{F}_1,\ldots,\mathbf{F}_K]$.

	At each user, linear receive combining is implemented using a single-layer MiLAC. Specifically, user~$k$ employs an $(N_k+D_k)$-port MiLAC. Let $T_k= N_k+D_k$, and denote the scattering matrices of the receiver-side MiLAC by
$\bm{\Psi}_k\in\mathbb{C}^{T_k\times T_k}$.
Following the models of MiLAC~1 and MiLAC~2, the analog combining matrix $\mathbf{U}_k\in\mathbb{C}^{N_k\times D_k}$ is given by
		\begin{equation}\label{eq:psi}
    \mathbf{U}_k
   =
    \frac{1}{2}
    [\bm{\Psi}_k]_{D_k+1:T_k,\,1:D_k},
    \qquad
    \forall k\in \mathcal{K},
\end{equation}
where $\mathcal{K} = \{1,\ldots,K\}$ is the user index set.

     Treating multi-user interference as noise, the estimate of the data symbol vector $\mathbf{s}_k$ is
\begin{equation}\label{eq:receive-signal}
    \hat{\mathbf{s}}_k
    =
    \mathbf{U}_k^{\mathsf T}\mathbf{y}_k. \footnote{The receive-side MiLAC is assumed to be lossless and therefore introduces no additional thermal noise. In practice, dissipative components in a lossy MiLAC may introduce insertion loss and additional thermal noise, which are not considered in the present model.}
\end{equation}
Define the effective transmit beamformer for user~$k$ as $\mathbf{G}_k
    =
    \mathbf{W}\mathbf{P}^{\frac{1}{2}}\mathbf{F}_k
    \in\mathbb{C}^{L\times D_k}$. Then, the achievable rate of user~$k$ is
\begin{equation}
    R_k
    =
    \log_2\det\left(
    \mathbf{I}_{D_k}
    +
    \mathbf{U}_k^{\mathsf T}
    \mathbf{H}_k
    \mathbf{G}_k\mathbf{G}_k^{\mathsf H}
    \mathbf{H}_k^{\mathsf H}
    \mathbf{U}_k^*
    \mathbf{J}_k^{-1}
    \right),
\end{equation}
where $\mathbf{J}_k
    =
    \sum_{i\neq k}
    \mathbf{U}_k^{\mathsf T}
    \mathbf{H}_k
    \mathbf{G}_i\mathbf{G}_i^{\mathsf H}
    \mathbf{H}_k^{\mathsf H}
    \mathbf{U}_k^*
    +
    4\sigma_k^2
    \mathbf{U}_k^{\mathsf T}\mathbf{U}_k^*$ denotes the interference-plus-noise covariance matrix that accounts for multi-user interference and noise contribution in the MIMO system.

	Since all MiLACs are assumed to be lossless and reciprocal, their
scattering matrices are unitary and symmetric, and therefore satisfy
\begin{subequations}\label{eq:scattering-constraints}
    \begin{align}
        \bm{\Theta}_m^{\mathsf H}\bm{\Theta}_m
        &=
        \mathbf{I}_{M_m},
        &
        \bm{\Theta}_m
        &=
        \bm{\Theta}_m^{\mathsf T},
        \quad
        \forall m\in\{1,2\},
        \label{eq:con-theta-m}\\
        \bm{\Psi}_k^{\mathsf H}\bm{\Psi}_k
        &=
        \mathbf{I}_{T_k},
        &
        \bm{\Psi}_k
        &=
        \bm{\Psi}_k^{\mathsf T},
        \quad
        \forall k\in\mathcal{K}.
        \label{eq:cons-psi}
    \end{align}
\end{subequations}

\subsection{Comparison Among Different Transceiver Architectures}

\begin{table*}[t!]
	\centering
	\caption{Comparison of Different Transceiver Architectures.}
	\label{tab:transceiver}
	\small
	\setlength{\tabcolsep}{5pt}
	\renewcommand{\arraystretch}{1.15}
	\begin{tabular}{ccccccc}
		\toprule
		\multicolumn{2}{c}{Architecture}
		& \multicolumn{2}{c}{Number of RF Chains}
		& \multicolumn{2}{c}{Converter Resolution}
		& \multirow{2}{*}{\shortstack{Fully Digital\\Sum-Rate Equivalence}} \\
		\cmidrule(lr){1-2}
		\cmidrule(lr){3-4}
		\cmidrule(lr){5-6}
		TX & RX & TX & RX & DAC & ADC & \\
		\midrule
		
		Fully digital
		& Fully digital
		& $L$ & $N_k$
		& High & High
		& \checkmark \\
		
		One-layer MiLAC
		& One-layer MiLAC
		& $D$ & $D_k$
		& Low & Low
		& $\times$ \\
		
		Hybrid digital-MiLAC 
		&  One-layer MiLAC
		& $D$ & $D_k$
		& High & Low
		& \checkmark \\
		
		Digital-PS hybrid
		& Digital-PS hybrid
		& $D$ & $D_k$
		& High & High
		& $\times$ \\
		
		\textbf{Two-layer MiLAC}
		& \textbf{One-layer MiLAC}
		& \textbf{$D$} & \textbf{$D_k$}
		& \textbf{Low} & \textbf{Low}
		& \textbf{\checkmark} \\
		
		\bottomrule
	\end{tabular}
\vspace{-0.4cm}
		\end{table*}

	Table~\ref{tab:transceiver} compares the hardware requirements and sum-rate capabilities of different transceiver architectures. In the conventional fully digital architecture, each	transmit or receive antenna is connected to a dedicated RF chain, resulting in $L$ RF chains at the transmitter and $N_k$ RF chains at user $k$, together with high-resolution DACs and ADCs. In contrast, the MiLAC-based architectures reduce the number of RF chains to the number of transmitted and received data streams, i.e., $D$ at the transmitter and $D_k$ at user $k$. The one-layer MiLAC transmitter and receiver architecture further enables	the use of low-resolution data converters, but its constrained analog beamforming capability prevents it from attaining the sum-rate performance of a fully digital transceiver. The hybrid digital-MiLAC architecture at the transmitter can recover the fully digital beamforming performance, but requires digital precoding and correspondingly higher-resolution data converters. Moreover, with the same number of high-resolution RF chains, digital-PS hybrid architecture still cannot recover the fully digital beamforming performance. By comparison, the proposed transceiver, consisting of a two-layer MiLAC transmitter and a	one-layer MiLAC receiver, simultaneously reduces the number of RF	chains and retains the potential to employ low-resolution data converters. As established in the next section, it can also attain the same maximum sum-rate as the fully digital transceiver under the considered system model. Therefore, the proposed architecture combines the beamforming flexibility of fully digital processing with the reduced number of RF chains and analog-processing advantages of MiLAC.

\section{Fully Digital Performance Equivalence of the Proposed Transceiver}\label{sec:optimality}
In this section, we rigorously prove that the proposed MiLAC-aided transceiver achieves fully digital performance under the considered system model. Specifically, we first show that the two-layer MiLAC architecture at the transmitter can exactly realize an arbitrary fully digital transmit beamformer.
We then demonstrate that a single-layer MiLAC architecture at each receiver can attain the same achievable rate as optimal fully digital linear receive combining.

\subsection{Fully Digital Equivalence at the Transmitter}\label{sec:optimality-tx}
In this subsection, we establish that the proposed two-layer MiLAC architecture at the transmitter can exactly realize an arbitrary fully digital transmit beamformer under the considered ideal MiLAC model. The main result is stated in Theorem~\ref{the:MiLAC-HBF}.

Before introducing the theorem, we first review the relationship of the data symbol vector and the transmit signal in a fully digital system \cite{nerini2025milac-capacity}:
\begin{equation}
	\mathbf{x}=\frac{1}{2}\mathbf{P}_d\mathbf{s},
\end{equation}
where $\mathbf{P}_d\in \mathbb{C}^{L \times D}$ is the digital beamforming matrix.
\begin{theorem}\label{the:MiLAC-HBF}
	Consider a fully digital beamformer $\mathbf{P}_d \in \mathbb{C}^{L \times D}$ for the considered multi-user MIMO system.  Let its compact singular value decomposition (SVD) be
	\begin{align}\label{eq:Pd-SVD}
		\mathbf{P}_d
    =
    \overline{\mathbf{U}}_1 \;
    \overline{\bm{\Sigma}}\;
    \overline{\mathbf{V}}^{\mathsf H},
	\end{align}
	where
$\overline{\mathbf{U}}_1\in\mathbb{C}^{L\times D}$ has orthonormal
columns, $\overline{\bm{\Sigma}}\in\mathbb{R}^{D\times D}$ is diagonal and
contains the singular values of $\mathbf{P}_d$, and
$\overline{\mathbf{V}}\in\mathbb{C}^{D\times D}$ is unitary. Let
$\overline{\mathbf{U}}
=
[\overline{\mathbf{U}}_1,\overline{\mathbf{U}}_2]
\in\mathbb{C}^{L\times L}$
be a unitary completion of $\overline{\mathbf{U}}_1$.

Then, under the ideal two-layer MiLAC model considered in Section~II, there always exists a feasible configuration of the two MiLAC layers and the intermediate $D$ amplifiers whose effective beamforming
matrix is exactly $\mathbf{P}_d$. One such construction is
\begin{equation}\label{eq:theta-star}
    \bm{\Theta}_1^\star
    =
    \begin{bmatrix}
        \mathbf{0}_{D\times D}
        &
        \overline{\mathbf{V}}^{\mathsf *}
        \\
        \overline{\mathbf{V}}^{\mathsf H}
        &
        \mathbf{0}_{D\times D}
    \end{bmatrix},
\end{equation}
and
\begin{equation}\label{eq:phi-star}
    \bm{\Theta}_2^\star
    =
    \begin{bmatrix}
        \mathbf{0}_{D\times D}
        &
        \overline{\mathbf{U}}_1^{\mathsf T}
        \\
        \overline{\mathbf{U}}_1
        &
        -\overline{\mathbf{U}}_2
        \overline{\mathbf{U}}_2^{\mathsf T}
    \end{bmatrix}.
\end{equation}
The corresponding scattering matrix for the amplifiers $\mathbf S_{A,21}^\star \in \mathbb{R}^{D\times D}$ and power-allocation matrix $\mathbf{P}^\star \in \mathbb{R}^{D\times D}$ are
given by
\begin{equation}\label{eq:g-star}
    \mathbf{S}_{A,21}^\star
    =
    \overline{\bm{\Sigma}},
\end{equation}
and
\begin{equation}\label{eq:p-star}
    (\mathbf{P}^\star)^{\frac{1}{2}}
    =
    \mathbf{S}_{A,21}^\star
    =
    \overline{\bm{\Sigma}}.
\end{equation}
Accordingly, the first- and second-stage analog beamforming matrices $\mathbf{F}^\star \in \mathbb{C}^{D \times D}, \mathbf{W}^\star \in \mathbb{C}^{L \times D}$ are
\begin{equation}
    \mathbf{F}^\star
    =
    [\bm{\Theta}_1^\star]_{D+1:2D,\,1:D}
    =
    \overline{\mathbf{V}}^{\mathsf H},
\end{equation}
and
\begin{equation}
    \mathbf{W}^\star
    =
    [\bm{\Theta}_2^\star]_{D+1:L+D,\,1:D}
    =
    \overline{\mathbf{U}}_1.
\end{equation}
The resulting effective transmit beamformer satisfies
\begin{equation}
    \mathbf{W}^\star
    (\mathbf{P}^\star)^{\frac{1}{2}}
    \mathbf{F}^\star
    =
    \mathbf{P}_d.
\end{equation}
Hence, the proposed two-layer MiLAC architecture can exactly reproduce any fully digital transmit beamformer under the considered ideal model.
\end{theorem}

\textit{Proof:}
The construction follows directly from the SVD in \eqref{eq:Pd-SVD}. Specifically, MiLAC~1 realizes the right singular directions $\overline{\mathbf{V}}^{\mathsf H}$, while MiLAC~2 realizes the left
singular directions $\overline{\mathbf{U}}_1$.
The intermediate $D$ amplifiers implement the corresponding singular
values through $\mathbf{S}_{A,21}^\star=\overline{\bm{\Sigma}}$.
The scattering matrices in \eqref{eq:theta-star} and \eqref{eq:phi-star} are symmetric and unitary, and are therefore
feasible under the adopted MiLAC scattering-matrix model. Consequently, $\mathbf{W}^\star
(\mathbf{P}^\star)^{\frac{1}{2}}
\mathbf{F}^\star
=
\overline{\mathbf{U}}_1
\overline{\bm{\Sigma}}\;
\overline{\mathbf{V}}^{\mathsf H}
=
\mathbf{P}_d$.
The detailed feasibility verification is provided in
Appendix~\ref{app:proof-theorem1}.
\hfill$\square$

Theorem~\ref{the:MiLAC-HBF} shows that the additional MiLAC layer
removes the beamforming restriction imposed by a
one-layer MiLAC transmitter. The result also provides a direct interpretation of the proposed architecture: the two MiLAC layers realize the left and right singular directions of the fully digital beamformer, while the intermediate amplifiers implement its singular values. Hence, under the considered ideal model, the proposed transmitter recovers the full beamforming flexibility of a fully digital architecture without introducing digital precoding.


\subsection{Fully Digital Equivalence at the Receiver}\label{sec:optimality-rx}
In this subsection, we show that a single-layer MiLAC architecture at each receiver can attain the same achievable rate as optimal fully digital linear receive combining. The main result is stated in Theorem~\ref{the:opt-1MiLAC}.

\begin{theorem}\label{the:opt-1MiLAC}
	Consider the downlink multi-user MIMO system
	\begin{equation}\label{eq:re-signal}
		\mathbf y_k
		= \frac{1}{2}
		\mathbf H_k
		\sum_{i=1}^{K}\mathbf G_i\mathbf s_i
		+
		\mathbf n_k,
		\qquad k\in\mathcal K,
	\end{equation}
	where 
	$\mathbf G_i\in\mathbb C^{L\times D_i}$ is the transmit precoder for user $i$, $\mathbf{H}_k\in\mathbb{C}^{N_k\times L}$ denotes the channel matrix from the transmitter to user~$k$ and
	$\mathbf n_k\sim
	\mathcal{CN}(\mathbf 0,\sigma_k^2\mathbf I_{N_k})$, with
	$\sigma_k^2>0$. Assume that
	$\operatorname{rank}(\mathbf H_k\mathbf G_k)=D_k$ for all
	$k\in\mathcal K$, and that multi-user interference is treated as noise.

    For user~$k$, define the set of receive combining matrices $\{\mathbf U_k\in\mathbb C^{N_k\times D_k}\}_{k=1}^K$ realizable by a fully connected, lossless, and reciprocal one-layer MiLAC as
\begin{equation}
\begin{split}
    \mathcal U_k^{\mathrm{MiLAC}}
    =
    \big\{\mathbf U_k=
    \frac{1}{2}
    [\bm{\Psi}_k]_{D_k+1:D_k+N_k,\,1:D_k}
    \; | \;\\
    \bm{\Psi}_k^{\mathsf H}\bm{\Psi}_k
    =
    \mathbf I_{N_k+D_k},
    \;
    \bm{\Psi}_k
    =
    \bm{\Psi}_k^{\mathsf T}
    \big\}.
\end{split}
\end{equation}
	Then, for any fixed set of transmit precoders $\{\mathbf G_i\}_{i=1}^{K}$, a single MiLAC layer at each receiver achieves the optimal rate attainable with unconstrained fully digital linear receive combining, i.e.,
\begin{equation}
    \max_{\mathbf U_k\in\mathcal U_k^{\mathrm{MiLAC}}}
    R_k
    =
    \max_{\mathbf U_k\in\mathbb C^{N_k\times D_k}}
    R_k,
    \qquad
    \forall k\in\mathcal K.
\end{equation}
Hence, under the stated assumptions, a single-layer MiLAC at each receiver is sufficient to attain the optimal fully digital linear-combining rate.
\end{theorem}

\textit{Proof:}
For a fixed set of transmit precoders, the optimal fully digital linear combining performance for user~$k$ can be attained by the MMSE combiner $\mathbf U_k^{\mathrm d}\in\mathbb C^{N_k\times D_k}$. A single-layer MiLAC can realize any combining matrix whose spectral norm does not exceed $1/2$ \cite{limits-milac}. Although $\mathbf U_k^{\mathrm d}$ may not directly satisfy this constraint, it can be scaled without affecting the achievable rate.

Specifically, define $\widetilde{\mathbf U}_k
    =
    c_k\left(\mathbf U_k^{\mathrm d}\right)^*$ and $c_k
    =
    \min\{
    1,\,
    \frac{1}{2\|\mathbf U_k^{\mathrm d}\|_2}
    \}$, such that $\|\widetilde{\mathbf U}_k\|_2\leq 1/2$. Since $c_k\neq 0$, this uniform scaling affects the desired signal, interference, and noise components equally, leaving the achievable rate unchanged.
Hence, $\widetilde{\mathbf U}_k$ satisfies the MiLAC spectral-norm constraint while achieving the same rate as $\mathbf U_k^{\mathrm d}$.

Using an SVD-based construction, we obtain a symmetric and unitary scattering matrix $\bm{\Psi}_k$ satisfying
\begin{equation}
    \frac{1}{2}
    [\bm{\Psi}_k]_{D_k+1:D_k+N_k,\,1:D_k}
    =
    \widetilde{\mathbf U}_k.
\end{equation}
Therefore, a single MiLAC layer can realize $\widetilde{\mathbf U}_k$ and achieve the same rate as the optimal fully digital linear receive combiner. The detailed construction and proof are provided in Appendix~\ref{app:proof-theorem2}.
\hfill$\square$

Theorem~\ref{the:opt-1MiLAC} highlights a fundamental difference between transmit and receive beamforming with a single-layer MiLAC. A single-layer MiLAC can realize the beamforming directions of an arbitrary linear beamformer after an appropriate scaling to satisfy its spectral-norm constraint. At the transmitter, such a scaling directly reduces the radiated signal power and therefore generally incurs a performance loss for a given transmit power constraint. At the receiver, in contrast, a nonzero scaling of the combining matrix scales the desired signal, interference, and noise by the same factor and hence leaves the achievable rate unchanged. Consequently, one layer of MiLAC is sufficient for rate-optimal receive combining, whereas an additional MiLAC layer is required at the transmitter to reproduce a fully digital beamformer without incurring this power loss.

\begin{remark}
	 A symmetric-unitary scattering matrix $\bm\Theta$ admits a finite-susceptance realization through the inverse Cayley transform if and only if	
	$
	\det(\mathbf I+\bm\Theta)\neq 0,
	$
	or equivalently, $-1\notin\operatorname{spec}(\bm\Theta)$. Thus, symmetry and unitarity alone do not guarantee finite-susceptance realizability. This additional condition can be satisfied without affecting the equivalence results. 	For the receive architecture in Theorem \ref{the:opt-1MiLAC}, consider the phase rotation
	$
	\widetilde{\bm\Psi}_k=e^{j\phi_k}\bm\Psi_k.
	$ This transformation preserves symmetry and unitarity. Since $\bm\Psi_k$ has finitely many eigenvalues, $\phi_k$ can always be chosen such that $-1\notin\operatorname{spec}(\widetilde{\bm\Psi}_k)$. The resulting combiner $\widetilde{\mathbf U}_k=e^{j\phi_k}\mathbf U_k$ achieves the same rate as $\mathbf U_k$. For the transmit architecture in Theorem \ref{the:MiLAC-HBF}, the phase freedom in the SVD-based construction can likewise be exploited to exclude $-1$ from the spectra of both scattering matrices while preserving the realized transmit beamformer.
\end{remark}

\section{Sum-Rate Optimization and Transceiver Design}\label{sec:pro-alg}
In this section, we first formulate the sum-rate maximization problem for the proposed MiLAC-aided transceiver based on the system model in Section~\ref{Sec:system model}. We then exploit the fully digital equivalence results established in Section~\ref{sec:optimality} to transform the MiLAC-aided transceiver design into an equivalent fully digital optimization problem, followed by a closed-form MiLAC realization.

\subsection{MiLAC-Aided Sum-Rate Maximization}
To characterize the maximum achievable sum rate of the proposed MiLAC-aided transceiver, we jointly optimize the transmitter- and receiver-side MiLAC configurations together with the intermediate power allocation. Specifically, we formulate the following sum-rate maximization problem subject to the power-allocation constraint and the structural constraints imposed by the MiLAC networks:
\begin{subequations}\label{eq:pro-formula}
    \begin{align}
        \max_{\bm{\Theta}_1,\,\bm{\Theta}_2,\,\mathbf{P},
        \{\bm{\Psi}_k\}_{k=1}^{K}}
        \quad
        & \sum_{k=1}^{K}
        R_k
        \left(
        \bm{\Theta}_1,\bm{\Theta}_2,\mathbf{P},
        \{\bm{\Psi}_i\}_{i=1}^{K}
        \right)
        \\
        \operatorname{s.t.}\quad
        & \operatorname{tr}(\mathbf{P})\leq P_t
        \label{eq:con-power},\\
        & \eqref{eq:scattering-constraints}.
        \notag
    \end{align}
\end{subequations}
Constraints~\eqref{eq:scattering-constraints} enforce the
symmetry and unitarity of the scattering matrices, thereby ensuring
consistency with the adopted lossless and reciprocal MiLAC model.
Constraint~\eqref{eq:con-power} imposes the power-allocation budget on the intermediate amplifier stage. 

Problem~\eqref{eq:pro-formula} is highly nonconvex because the optimization variables are coupled through the achievable-rate expressions, while the scattering matrices are subject to nonconvex symmetry and unitarity constraints.
Rather than directly solving \eqref{eq:pro-formula}, we exploit the equivalence theorems established in Section~\ref{sec:optimality} to derive a fully digital formulation, as described in the next subsection.
%

\subsection{Equivalent Fully Digital Optimization and MiLAC Realization}
Building on the equivalence theorems established in Section~\ref{sec:optimality}, we propose the following two-step approach:
\begin{itemize}
	\item \textbf{Step 1:} Solve the sum-rate maximization problem for a conventional fully digital transceiver to obtain the corresponding transmit beamforming matrix $\mathbf{P}_{\mathrm{d}}^\star\in\mathbb{C}^{L\times D}$ and the MMSE receive combining matrices $\{\mathbf{U}_{k}^{\mathrm{d},\star}\}_{k=1}^{K}$.
	
	\item \textbf{Step 2:} Given $\mathbf{P}_{\mathrm{d}}^\star$, apply Theorem~\ref{the:MiLAC-HBF} to construct the transmitter-side scattering matrices
    $\bm{\Theta}_1^\star$ and $\bm{\Theta}_2^\star$, together with the intermediate power-allocation matrix $\mathbf{P}^\star$. Similarly, given $\{\mathbf{U}_{k}^{\mathrm d,\star}\}_{k=1}^{K}$, apply Theorem~\ref{the:opt-1MiLAC} to construct a feasible receiver-side scattering matrix $\bm{\Psi}_k^\star$ for each user.
\end{itemize}
Since the closed-form MiLAC realizations in Step~2 are provided by Theorems~\ref{the:MiLAC-HBF} and~\ref{the:opt-1MiLAC}, the remaining task is to determine the fully digital transmit beamformer and the corresponding receive combiners in Step~1.

	For a conventional multi-user MIMO system, let $\mathbf{P}_{\mathrm d}
    =
    [\mathbf{P}_1,\mathbf{P}_2,\ldots,\mathbf{P}_K]
    \in\mathbb{C}^{L\times D}$, where
$\mathbf{P}_k\in\mathbb{C}^{L\times D_k}$
denotes the transmit beamformer for user~$k$.
The fully digital sum-rate maximization problem is formulated as 
	\begin{subequations}\label{eq:pro-formula-di}
		\begin{align}
			\max_{\mathbf{P}_{\mathrm d},
        \{\mathbf{U}_k^{\mathrm d}\}_{k=1}^{K}}
        \quad
        & \sum_{k=1}^{K}R_k^{\mathrm d}
        \\
        \operatorname{s.t.}\quad
        & \sum_{k=1}^{K} \operatorname{tr} \left( \mathbf{P}_k\mathbf{P}_k^{\mathsf H}
        \right)
        \leq P_t,
 \label{eq:power-digital}
		\end{align}
	\end{subequations}
	where 
	\begin{equation}\label{eq:rate-di}
		R_k^d = \log_2 \det \left( \mathbf{I} + 	{\mathbf{U}_k^d}^\mathsf{H}  \mathbf{H}_k\mathbf{P}_k\mathbf{P}_k^\mathsf{H}\mathbf{H}_k^\mathsf{H}\mathbf{U}_k^d\tilde{\mathbf{J}}_k^{-1} \right),
	\end{equation}
	and $\tilde{\mathbf{J}}_k=\sum_{i \neq k}{\mathbf{U}_k^d}^\mathsf{H}\mathbf{H}_k\mathbf{P}_i\mathbf{P}_i^\mathsf{H}\mathbf{H}_k^\mathsf{H}\mathbf{U}_k^d+4\sigma_k^2{\mathbf{U}_k^d}^\mathsf{H} \mathbf{U}_k^d$. 
    Since the MiLACs are assumed to be lossless, they do not introduce any power dissipation. Accordingly, the proposed MiLAC transceiver and its fully digital counterpart are subject to the same radiated transmit-power budget, with the internal power-allocation variables governing the power distribution across the signal streams.
    
    Although problem~\eqref{eq:pro-formula-di} is nonconvex, it can be efficiently addressed using established alternating-optimization methods, such as fractional programming (FP) \cite{shen2018fractional} and weighted minimum mean-square error (WMMSE) optimization \cite{WMMSE2011,rethingkingWMMSE}. In this work, we adopt WMMSE to obtain a stationary solution $\mathbf{P}_{\mathrm d}^{\star}$ and the corresponding MMSE combiners $\{\mathbf{U}_{k}^{\mathrm d,\star}\}_{k=1}^{K}$.
The closed-form constructions in Theorems~\ref{the:MiLAC-HBF} and~\ref{the:opt-1MiLAC} are then applied to realize the same transmit beamformer and receive-combining performance using the proposed MiLAC-aided architecture.

\section{Simulation}\label{sec:sim}

In this section, we evaluate the sum-rate performance of the proposed transceiver architecture incorporating a two-layer MiLAC transmitter, using the algorithm developed in Section~\ref{sec:pro-alg}. We compare its performance with baseline architectures to validate the optimality results established in Section~\ref{sec:optimality}.

We consider independent Rayleigh fading channels, where the entries of each channel matrix
$\mathbf H_k\in\mathbb C^{N_k\times L}$ for user-$k$ are independently drawn from $\mathcal{CN}(0,1)$. The noise variance is normalized to $\sigma_k^2=1$ for each user, and the transmit SNR is defined as $\mathrm{SNR}
	=
	10\log_{10}(\frac{P_t}{\sigma_k^2})\ \mathrm{dB}$.
The relative sum-rate performance tolerance is set to $10^{-5}$, and the iterative algorithms terminate when this criterion is satisfied. The results are averaged over $100$ independent channel realizations.

The following five transceiver architectures are compared:

\begin{enumerate}
	\item \emph{Fully digital}: Both the transmit precoder and receive combiners are implemented digitally and jointly optimized using WMMSE, with closed-form MMSE receive-combiner updates \cite{WMMSE2011}. 
	
	\item \emph{MiLAC HBF}: This architecture employs hybrid digital-MiLAC beamforming at the transmitter and one-layer MiLAC combining at each receiver. The fully digital transmit precoder is factorized via SVD and exactly reconstructed through cascaded digital and MiLAC processing stages~\cite{limits-milac}.
	
	\item \emph{One-layer MiLAC}: One-layer MiLAC architecture implements beamforming at the transmitter and each receiver. The transmit and receive beamforming matrices, together with the transmit power allocation, are jointly optimized using the WMMSE-PGD algorithm \cite{limits-milac} under the corresponding spectral-norm and power constraints. 
	
	\item \emph{PS HBF}: Phase-shifter-based hybrid architecture implements beamforming at the transmitter and receivers, using $D$ transmit RF chains and $D_k$ receive RF chains at each receiver. The transmit beamformer is updated using PSLA-based algorithm \cite{two-layer-con} under constant-modulus constraints on its RF matrix. Each receive RF matrix is obtained by elementwise phase projection of the corresponding unconstrained MMSE combiner, after which the digital MMSE combiner is recomputed. The transmit and receive updates are performed alternately.
	
	\item \emph{Proposed transceiver}: The transmitter employs two-layer MiLAC architecture, while each receiver employs one-layer MiLAC architecture. The fully digital transmit precoder is exactly reproduced through the SVD-based construction for two-layer MiLAC architecture. At user-$k$, the fully digital MMSE combiner is uniformly scaled to satisfy the one-layer MiLAC spectral-norm constraint.
\end{enumerate}

\begin{figure}[t!]
	\centering
	\subfloat[\label{fig:tx-rx}Comparison of transceiver architectures.]
	{\includegraphics[width=0.8\columnwidth]{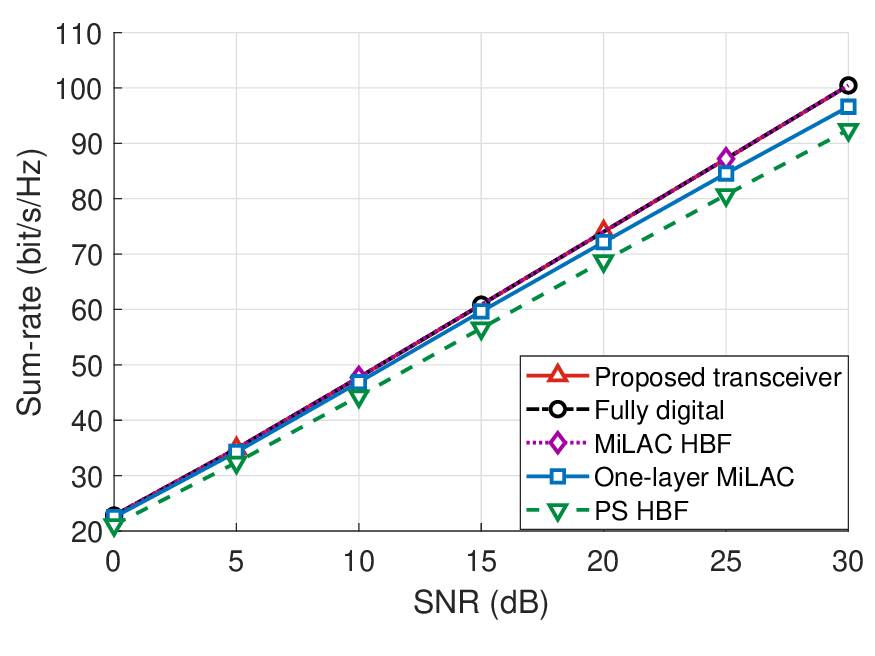}}
	\hfill
	\subfloat[\label{fig:tx}Comparison of  transmitter architectures.]
	{\includegraphics[width=0.8\columnwidth]{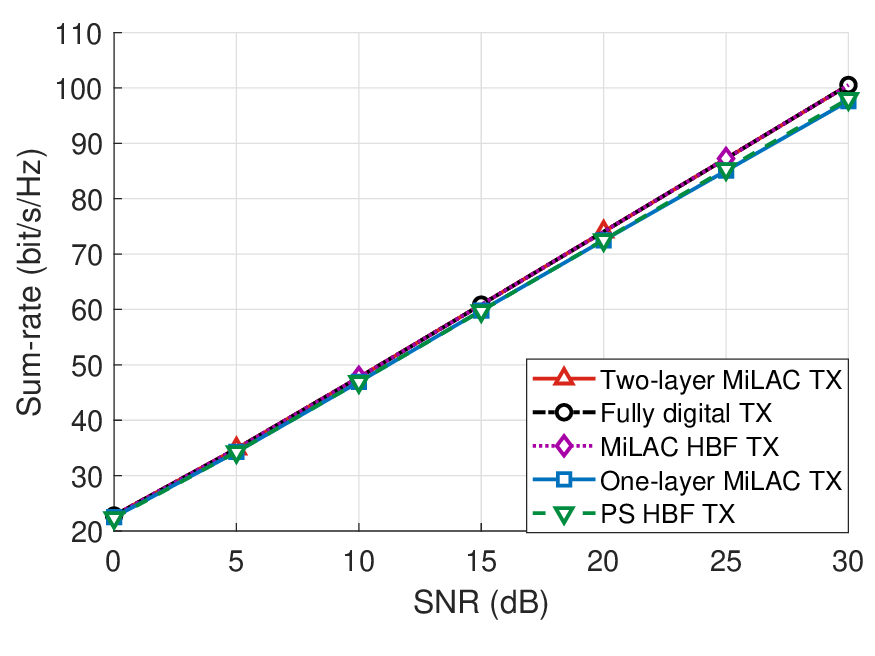}}
	\hfill
	\subfloat[\label{fig:rx}Comparison of  receiver architectures.]
	{\includegraphics[width=0.8\columnwidth]{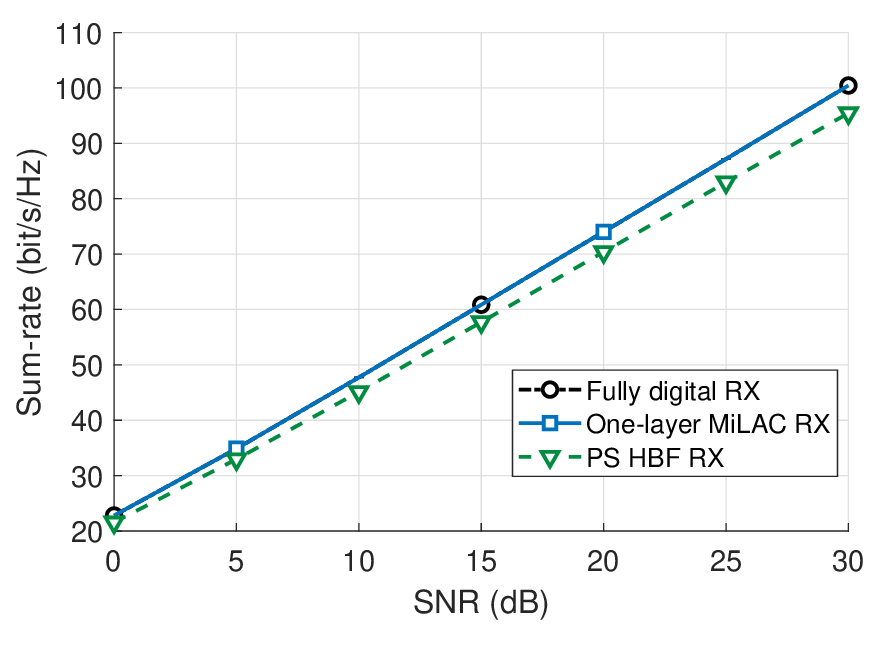}}
	\caption{Sum-rate versus SNR for $L=32$,
		$K=4$, $N_k=8$, and $D_k=2$.}
	\label{fig:architecture_snr}
	\vspace{-0.5cm}
\end{figure}

Fig.~\ref{fig:tx-rx} compares the achievable sum-rate versus transmit SNR for $L=32$, $K=4$, $N_k=8$, and $D_k=2$. The \emph{proposed transceiver} and the \emph{MiLAC HBF} architecture match the \emph{fully digital} benchmark across the considered SNR range. In contrast, the \emph{one-layer MiLAC} and \emph{PS HBF} architectures exhibit performance gaps that widen at high SNR. To identify the sources of these gaps, Fig.~\ref{fig:tx} compares transmitter architectures with fully digital receivers, while Fig.~\ref{fig:rx} compares receiver architectures with a fully digital transmitter. The transmit precoder and receive combiners are jointly optimized for each configuration. With fully digital receivers, the two-layer MiLAC and MiLAC-aided hybrid transmitters match the fully digital benchmark, whereas the one-layer MiLAC and phase-shifter-based hybrid transmitters exhibit residual gaps. With a fully digital transmitter, one-layer MiLAC receivers match fully digital combining, while phase-shifter-based hybrid receivers incur a noticeable loss. These results indicate that, under the considered model, the performance limitation of the \emph{one-layer MiLAC} architecture originates at the transmitter, with no observable receive-side penalty. The \emph{PS HBF} architecture exhibits gaps at both sides, with a larger gap in the receiver-side comparison for this configuration. These findings support the proposed combination of a two-layer MiLAC transmitter and one-layer MiLAC receivers.

\begin{figure}[t]
	\centering
	\includegraphics[width=0.8\columnwidth]
	{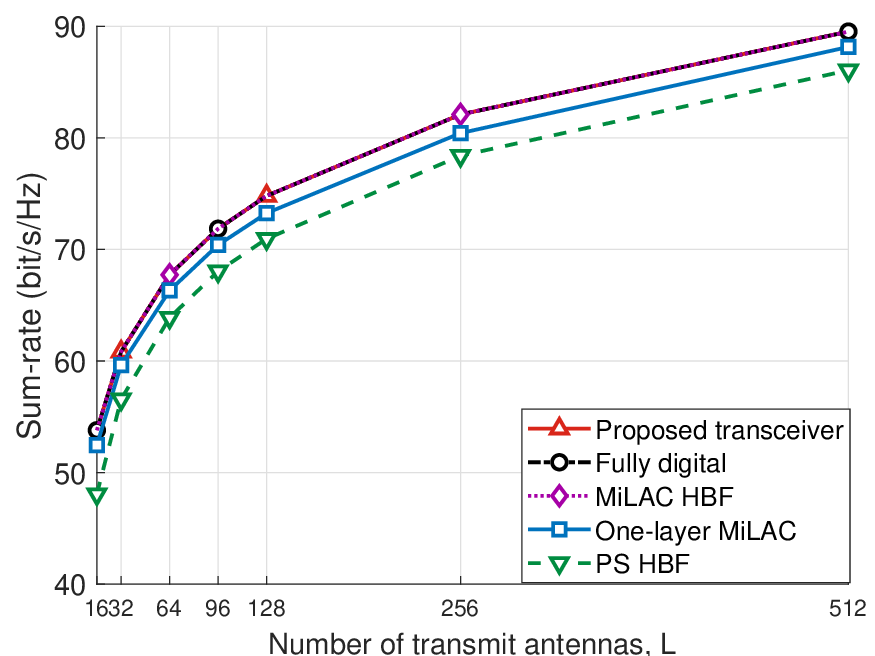}
	\caption{Sum-rate versus the number of transmit
		antennas for $K=4$, $N_k=8$, $D_k=2$, and
		$\mathrm{SNR}=15$ dB. }
	\label{fig:architecture_antennas}
\end{figure}

Fig.~\ref{fig:architecture_antennas} compares the achievable sum-rate versus the number of transmit antennas for $K=4$, $N_k=8$, $D_k=2$, and an SNR of $15$~dB, with $L$ ranging from $16$ to $512$. All architectures benefit from increasing $L$, owing to the greater array gain and additional spatial degrees of freedom. The \textit{MiLAC HBF} and \textit{proposed transceiver} architectures match the \textit{fully digital} benchmark throughout the considered range, whereas the \textit{one-layer MiLAC} and \textit{PS HBF} architectures exhibit residual performance gaps. For the \textit{one-layer MiLAC} architecture, the relative gap is smaller at large $L$, suggesting that additional spatial degrees of freedom help mitigate its transmit-side constraints.

\begin{figure}[t]
	\centering
	\includegraphics[width=0.8\columnwidth]
	{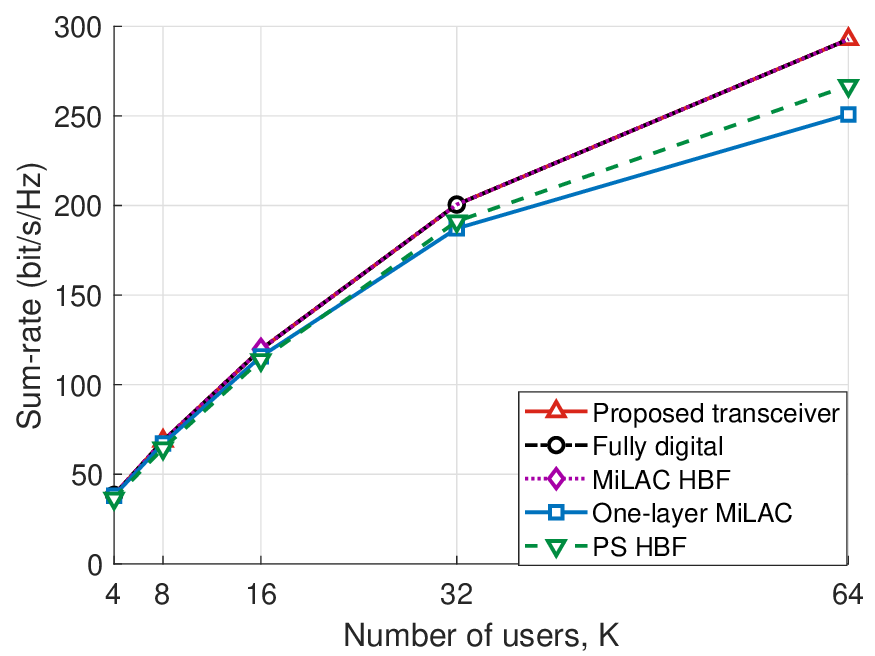}
	\caption{Sum-rate versus the number of users for
		$L=64$, $N_k=8$, $D_k=1$, and
		$\mathrm{SNR}=15$ dB. }
	\label{fig:architecture_users}
	\vspace{-0.5cm}
\end{figure}

Fig.~\ref{fig:architecture_users} compares the achievable sum-rate versus the number of users for $L=64$, $N_k=8$, $D_k=1$, and an SNR of $15$~dB. We vary $K\in\{4,8,16,32,64\}$, yielding $D=K\leq L$. The \textit{proposed transceiver} and \textit{MiLAC HBF} architectures match the fully digital benchmark across all considered values of $K$, confirming that the proposed architecture maintains fully digital performance as the number of simultaneously served users increases. The performance gap of the \textit{one-layer MiLAC} architecture widens with $K$. As the spatial load $D/L$ increases, fewer transmit dimensions are available per stream, making multi-user interference suppression more demanding. The constraints imposed by a single MiLAC limit the precoder's ability to jointly control stream directions and power allocation. For \textit{PS HBF}, the number of transmit RF chains equals $K=D$, so increasing $K$ also provides additional degrees of freedom in the digital and analog precoders. This increased flexibility helps \textit{PS HBF} outperform the \textit{one-layer MiLAC} architecture at larger $K$, although a gap to the fully digital benchmark remains.

	\section{Conclusion}\label{sec:conclusion}
In this paper, we proposed a two-layer MiLAC architecture to overcome the performance limitations of one-layer MiLAC transmitters in downlink multi-user MIMO networks. Using multiport microwave network theory, we established a general physics-compliant model that captures component interactions and derived a simplified model for unilateral and perfectly matched amplifiers. Under this simplified model with lossless and reciprocal MiLACs, we proved that the two-layer transmitter can realize any fully digital transmit beamformer, while a single-layer MiLAC receiver can achieve the same rate as fully digital linear combining. We derived closed-form mappings to the corresponding MiLAC implementations and developed a sum-rate maximization framework to jointly design the power allocation and scattering matrices. Numerical results confirmed that the proposed transceiver matches the sum-rate performance of its fully digital counterpart under the considered system model. These findings establish the potential of combining a two-layer MiLAC transmitter with one-layer MiLAC receivers to realize fully digital beamforming performance through analog-domain processing.

    	\begin{appendices}
        \section{Proof for Theorem \ref{the:MiLAC-HBF}}\label{app:proof-theorem1}

Following the notation in \eqref{eq:Theta-partition}, let $M_1=2D$ and $M_2=L+D$. For the two-layer MiLAC architecture proposed in Section \ref{sec:proposed-architecture}, we obtain its effective beamformer $ \mathbf{G} \in \mathbb{C}^{L \times D}$ at the transmitter as 
\begin{equation}\label{eq:eff-G}
	\mathbf{G}=\mathbf{W}\mathbf{P}^{\frac{1}{2}}\mathbf{F} = \bm \Theta_{2,21}\mathbf{S}_{A,21}\bm\Theta_{1,21}.
\end{equation}
The second equality in \eqref{eq:eff-G} is obtained by substituting \eqref{eq:theta1}, \eqref{eq:theta2} and \eqref{eq:amplifier-gain} into \eqref{eq:eff-G}. To prove Theorem~\ref{the:MiLAC-HBF}, it is sufficient to show that there exists a matrix $\mathbf{G}$ satisfying the constraints of the MiLAC architecture and is equivalent to the fully digital beamformer 
$\mathbf{P}_d$. This can be established by demonstrating that the following optimization problem always admits a global solution with an objective value of zero:
\begin{subequations}\label{eq:ls}
	\begin{align}
		\min_{\bm\Theta_1,\bm \Theta_2,\mathbf{P}^{\frac{1}{2}}} \quad& 
		\left\|\mathbf{P}_d-\bm\Theta_{2,21}\mathbf{S}_{A,21}\bm\Theta_{1,21}\right\|_\mathsf{F}^2 \label{eq:ls-obj}\\
		\text{s.t.} \quad 
		& \eqref{eq:amplifier-gain},\eqref{eq:con-theta-m},\eqref{eq:con-power}.
	\end{align}
\end{subequations}
To solve problem \eqref{eq:ls}, we first consider the SVD of $\mathbf{P}_d$ in \eqref{eq:Pd-SVD}. Since columns of $\mathbf{P}_d$ lie in $\operatorname{span}(\overline{\mathbf{U}}_1)$, it is natural to choose $\bm \Theta_{2,21}^\star = \overline{\mathbf{U}}_1$. For a fixed $\bm \Theta_{2,21}^\star$,  the objective in \eqref{eq:ls-obj} is minimized by choosing $\bm\Theta_{1,21}^\star = \overline{\mathbf{V}}^\mathsf{H}$. 
However, $\bm \Theta_{2,21}^\star$ and $\bm\Theta_{1,21}^\star$ are only sub-blocks of $\bm \Theta_2^\star$ and $\bm\Theta_1^\star$, respectively. We must therefore construct full matrices $\bm \Theta_2^\star$ and $\bm\Theta_1^\star$ that satisfy the constraints \eqref{eq:con-theta-m} in \eqref{eq:ls}. To this end, we begin by verifying the feasibility of $\bm \Theta_2$, followed by that of $\bm \Theta_1$. To ensure that the structural constraints in \eqref{eq:theta2} and \eqref{eq:con-theta-m} are satisfied, the feasibility of $\bm \Theta_2$ can be assessed by solving the following problem:
\begin{subequations}\label{eq:feasible-check}
	\begin{align}
		\text{find}\quad &\bm \Theta_{2,11},\bm \Theta_{2,22}\\
		\text{s.t.}\quad &
		\begin{bmatrix}
			\bm \Theta_{2,11} & \overline{\mathbf{U}}_1^\mathsf{T}\\
			\overline{\mathbf{U}}_1 & \bm \Theta_{2,22}
		\end{bmatrix}^\mathsf{H}
		\begin{bmatrix}
			\bm \Theta_{2,11} & \overline{\mathbf{U}}_1^\mathsf{T}\\
			\overline{\mathbf{U}}_1 & \bm \Theta_{2,22}
		\end{bmatrix}
		=\mathbf{I}_{L+D}, \label{eq:unit-norm}\\
		&
		\bm \Theta_{2,11}=\bm \Theta_{2,11}^\mathsf{T},\\
		&
		\bm \Theta_{2,22}=\bm \Theta_{2,22}^\mathsf{T}.
	\end{align}
\end{subequations}

Since $\overline{\mathbf{U}}_1^\mathsf{H}\overline{\mathbf{U}}_1=\mathbf{I}_D$, the first $D$ columns of $\bm \Theta_2$ must have unit $\ell_2$ norm. This forces $\bm \Theta_{2,11}^\star = \mathbf{0}_{D\times D}$.
The remaining feasibility conditions in \eqref{eq:feasible-check} reduce to
\begin{subequations}\label{eq:phi22}
	\begin{align}
		\overline{\mathbf{U}}_1^\mathsf{H} \bm \Theta_{2,22} &= \mathbf{0}_{D\times L},\\
		\overline{\mathbf{U}}_1^*\overline{\mathbf{U}}_1^\mathsf{T} + \bm \Theta_{2,22}^\mathsf{H}\bm \Theta_{2,22} &= \mathbf{I}_L,\\
		\bm \Theta_{2,22} &= \bm \Theta_{2,22}^\mathsf{T}.
	\end{align}
\end{subequations}
A valid solution for $\bm \Theta_{2,22}$ is $\bm \Theta_{2,22}^\star = -\overline{\mathbf{U}}_2\overline{\mathbf{U}}_2^\mathsf{T}$,
which satisfies all three constraints in \eqref{eq:phi22}. Notably, the selection of $\bm \Theta_{2,22}$ has no impact on system performance. Accordingly, the optimal solution for MiLAC 2 is given by \eqref{eq:phi-star}.
Following the same procedure as that used to obtain $\bm \Theta_2^\star$ in \eqref{eq:feasible-check}-\eqref{eq:phi22}, we obtain the optimal solution for the scattering matrix $\bm \Theta_1$ of MiLAC 1 as \eqref{eq:theta-star}.
\par
Now, it remains to optimize $\mathbf{P}^{\frac{1}{2}}$ in \eqref{eq:ls}.
With $\bm \Theta_{2,21}^\star=\overline{\mathbf{U}}_1$ and $\bm\Theta_{1,21}^\star=\overline{\mathbf{V}}^\mathsf{H}$, problem \eqref{eq:ls} reduces to
\begin{subequations}\label{eq:ls-P}
	\begin{align}
		\min_{\mathbf{P}^{\frac{1}{2}}} \quad &
		\big\|\overline{\bm{\Sigma}} - \mathbf{S}_{A,21}\big\|_\mathsf{F}^2\\
		\text{s.t.}\quad&
		\eqref{eq:amplifier-gain},\eqref{eq:con-power}.
	\end{align}
\end{subequations}
Notice that \eqref{eq:ls-P} is a constrained least-squares problem. 
	Since the given fully digital beamformer satisfies the power constraint, $\overline{\bm\Sigma}$ is
	feasible for problem \eqref{eq:ls-P}. Therefore, the global minimum is zero, attained at	$\mathbf S_{A,21}^{\star}=\overline{\bm\Sigma}$,
	which gives \eqref{eq:p-star}.

In summary, under the solution $(\bm\Theta_1^\star,\bm\Theta_2^\star,(\mathbf{P}^\star)^{\frac{1}{2}})$, the objective value of \eqref{eq:ls} is zero, and this solution satisfies
\begin{equation}
	\bm\Theta^\star_{2,21}\mathbf{S}_{A,21}^\star\bm\Theta^\star_{1,21}
	= \overline{\mathbf{U}}_1\overline{\bm{\Sigma}}\;\overline{\mathbf{V}}^\mathsf{H}
	= \mathbf{P}_d,
\end{equation}
which proves that two-layer MiLAC-aided beamforming can always reproduce the fully digital beamformer when both MiLAC 1 and MiLAC 2 are lossless and reciprocal. Therefore, we complete the proof.
\hfill$\square$

\section{Proof for Theorem \ref{the:opt-1MiLAC}}\label{app:proof-theorem2}

For analytical convenience, we normalize the received signal in \eqref{eq:re-signal} as
$\overline{\mathbf y}_k=2\mathbf y_k
=\sum_{i=1}^{K}\mathbf H_k\mathbf G_i\mathbf s_i
+\overline{\mathbf n}_k$,
where $\overline{\mathbf n}_k=2\mathbf n_k
\sim\mathcal{CN}(\mathbf 0,4\sigma_k^2\mathbf I_{N_k})$.
Since this nonzero scaling does not affect the
achievable rate, we use $\overline{\mathbf y}_k$
for the following MMSE analysis.
For user $k$, define the received-signal covariance matrix and the
	interference-plus-noise covariance matrix, respectively, as
	\begin{align}
		\mathbf R_{y,k}
		&=
		\sum_{i=1}^{K}
		\mathbf H_k\mathbf G_i\mathbf G_i^{\mathsf H}
		\mathbf H_k^{\mathsf H}
		+4\sigma_k^2\mathbf I_{N_k},
		\label{eq:received_covariance}\\
		\mathbf R_{-k}
		&=
		\sum_{\substack{i=1\\i\neq k}}^{K}
		\mathbf H_k\mathbf G_i\mathbf G_i^{\mathsf H}
		\mathbf H_k^{\mathsf H}
		+4\sigma_k^2\mathbf I_{N_k}.
		\label{eq:interference_covariance}
	\end{align}
	Accordingly, $\mathbf R_{y,k}
	=
	\mathbf R_{-k}
	+
	\mathbf H_k\mathbf G_k\mathbf G_k^{\mathsf H}
	\mathbf H_k^{\mathsf H}$. Since $\sigma_k^2>0$, both $\mathbf R_{y,k}$ and
	$\mathbf R_{-k}$ are positive definite.
	
	For a fully digital receive combining matrix
	$\mathbf U_k^{\mathrm d}\in\mathbb C^{N_k\times D_k}$, the estimate
	of $\mathbf s_k$ is given by $\widehat{\mathbf s}_k
		=
		(\mathbf U_k^{\mathrm d})^{\mathsf H}\overline{\mathbf y}_k$. The corresponding MSE matrix is
	\begin{align}
		\mathbf E_k
		={}&
		\mathbb E\left[
		(\widehat{\mathbf s}_k-\mathbf s_k)
		(\widehat{\mathbf s}_k-\mathbf s_k)^{\mathsf H}
		\right]                                                        \notag\\
		={}&
		\mathbf I_{D_k}
		-(\mathbf U_k^{\mathrm d})^{\mathsf H}\mathbf H_k\mathbf G_k
		-\mathbf G_k^{\mathsf H}\mathbf H_k^{\mathsf H}
		\mathbf U_k^{\mathrm d} +
		(\mathbf U_k^{\mathrm d})^{\mathsf H}
		\mathbf R_{y,k}\mathbf U_k^{\mathrm d}.
		\label{eq:rx_mse_matrix}
	\end{align}
	Differentiating $\operatorname{tr}(\mathbf E_k)$ with respect to
	$(\mathbf U_k^{\mathrm d})^*$ gives
	\begin{equation}
		\frac{\partial\operatorname{tr}(\mathbf E_k)}
		{\partial(\mathbf U_k^{\mathrm d})^*}
		=
		\mathbf R_{y,k}\mathbf U_k^{\mathrm d}
		-\mathbf H_k\mathbf G_k.
	\end{equation}
	Therefore, the unique linear MMSE combiner is
	\begin{equation}\label{eq:fd_mmse_receiver}
		\mathbf U_{k,\mathrm{MMSE}}^{\mathrm d}
		=
		\mathbf R_{y,k}^{-1}\mathbf H_k\mathbf G_k.
	\end{equation}
	
	Substituting \eqref{eq:fd_mmse_receiver} into
	\eqref{eq:rx_mse_matrix} yields
	\begin{align}
		\mathbf E_{k,\mathrm{MMSE}}
		&=
		\mathbf I_{D_k}
		-
		\mathbf G_k^{\mathsf H}\mathbf H_k^{\mathsf H}
		\mathbf R_{y,k}^{-1}\mathbf H_k\mathbf G_k                       \\
		&=
		\left(
		\mathbf I_{D_k}
		+
		\mathbf G_k^{\mathsf H}\mathbf H_k^{\mathsf H}
		\mathbf R_{-k}^{-1}\mathbf H_k\mathbf G_k
		\right)^{-1},
		\label{eq:mmse_error_rate_relation}
	\end{align}
	where the second equality follows from the matrix inversion
	lemma. Consequently, the rate attained by the MMSE combiner is
	\begin{align}
		R_k^{\mathrm d}
		&=
		-\log_2\det(\mathbf E_{k,\mathrm{MMSE}})                         \\
		&=
		\log_2\det\left(
		\mathbf I_{D_k}
		+
		\mathbf G_k^{\mathsf H}\mathbf H_k^{\mathsf H}
		\mathbf R_{-k}^{-1}\mathbf H_k\mathbf G_k
		\right).
		\label{eq:fd_mmse_rate}
	\end{align}
	
	To verify that this is the maximum rate attainable by a linear
	fully digital combiner, the matrix inversion lemma also gives
	\begin{align}
		\mathbf U_{k,\mathrm{MMSE}}^{\mathrm d}
		={}&
		\mathbf R_{-k}^{-1}\mathbf H_k\mathbf G_k
		\left(
		\mathbf I_{D_k}
		+
		\mathbf G_k^{\mathsf H}\mathbf H_k^{\mathsf H}
		\mathbf R_{-k}^{-1}\mathbf H_k\mathbf G_k
		\right)^{-1}.
		\label{eq:mmse_rate_optimal_form}
	\end{align}
	The rightmost matrix in \eqref{eq:mmse_rate_optimal_form} is
	nonsingular. Hence, the MMSE combiner and
	$\mathbf R_{-k}^{-1}\mathbf H_k\mathbf G_k$ span the same
	receive-combining subspace and differ only by a nonsingular
	right-side transformation. Since such a transformation does not
	change the achievable rate, \eqref{eq:fd_mmse_rate} is the maximum
	rate attainable by a fully digital linear combiner.
	
	Next, define the positive scaling factor
	\begin{equation}
		c_k
		=
		\min\left\{
		1,\,
		\frac{1}
		{2\|\mathbf U_{k,\mathrm{MMSE}}^{\mathrm d}\|_2}
		\right\}
		\label{eq:rx_scaling_factor}
	\end{equation}
	and the scaled and conjugated MMSE combiner
	\begin{equation}
		\widetilde{\mathbf U}_k
		=
		c_k\left(\mathbf U_{k,\mathrm{MMSE}}^{\mathrm d}\right)^*.
		\label{eq:scaled_mmse_receiver}
	\end{equation}
	The assumption
	$\operatorname{rank}(\mathbf H_k\mathbf G_k)=D_k$ ensures that
	$\|\mathbf U_{k,\mathrm{MMSE}}^{\mathrm d}\|_2>0$ and, hence,
	$c_k>0$. By construction,
	\begin{equation}\label{eq:rx_contraction}
		\|\widetilde{\mathbf U}_k\|_2
		\leq
		\frac{1}{2}.
	\end{equation}
	
	We next show that any matrix satisfying
	\eqref{eq:rx_contraction} can be implemented by a single-layer MiLAC.
	Consider the full SVD
	\begin{equation}
		2\widetilde{\mathbf U}_k
		=
		\overline{\mathbf Q}_k\boldsymbol{\Sigma}_k\overline{\mathbf P}_k^{\mathsf H},
		\label{eq:scaled_combiner_svd}
	\end{equation}
	where $\overline{\mathbf Q}_k\in\mathbb C^{N_k\times N_k}$ and
	$\overline{\mathbf P}_k\in\mathbb C^{D_k\times D_k}$ are unitary, while
	$\boldsymbol{\Sigma}_k\in\mathbb R^{N_k\times D_k}$ is rectangular
	diagonal. It follows from \eqref{eq:rx_contraction} that
	$\|\boldsymbol{\Sigma}_k\|_2\leq1$.
	
	Define
	\begin{align}
		\mathbf C_{D,k}
		&=
		\left(
		\mathbf I_{D_k}
		-\boldsymbol{\Sigma}_k^{\mathsf T}
		\boldsymbol{\Sigma}_k
		\right)^{1/2},                                                   \\
		\mathbf C_{N,k}
		&=
		\left(
		\mathbf I_{N_k}
		-\boldsymbol{\Sigma}_k
		\boldsymbol{\Sigma}_k^{\mathsf T}
		\right)^{1/2}.
	\end{align}
	These matrices are well defined because all singular values in
	$\boldsymbol{\Sigma}_k$ are no greater than one. Now construct
	\begin{equation}
		\boldsymbol{\Psi}_{0,k}
		=
		\begin{bmatrix}
			\mathbf C_{D,k}
			&
			\boldsymbol{\Sigma}_k^{\mathsf T}
			\\
			\boldsymbol{\Sigma}_k
			&
			-\mathbf C_{N,k}
		\end{bmatrix}.
		\label{eq:canonical_unitary_completion}
	\end{equation}
	Clearly,
	$\boldsymbol{\Psi}_{0,k}
	=\boldsymbol{\Psi}_{0,k}^{\mathsf T}$. Moreover, the definitions
	of $\mathbf C_{D,k}$ and $\mathbf C_{N,k}$ imply
	\begin{equation}
		\mathbf C_{N,k}\boldsymbol{\Sigma}_k
		=
		\boldsymbol{\Sigma}_k\mathbf C_{D,k}.
	\end{equation}
	It then follows by direct block multiplication that
	\begin{equation}
		\boldsymbol{\Psi}_{0,k}^{\mathsf H}
		\boldsymbol{\Psi}_{0,k}
		=
		\mathbf I_{N_k+D_k}.
	\end{equation}
	Thus, $\boldsymbol{\Psi}_{0,k}$ is both symmetric and unitary.
	
	Let
	\begin{equation}
		\mathbf T_k
		=
		\operatorname{blkdiag}
		\left(\overline{\mathbf P}_k^*,\overline{\mathbf Q}_k\right)
	\end{equation}
	and define
	\begin{equation}
		\boldsymbol{\Psi}_k
		=
		\mathbf T_k
		\boldsymbol{\Psi}_{0,k}
		\mathbf T_k^{\mathsf T}.
		\label{eq:rx_scattering_completion}
	\end{equation}
	Since $\mathbf T_k$ is unitary, the unitary congruence
	transformation in \eqref{eq:rx_scattering_completion} preserves
	both symmetry and unitarity. Therefore,
	\begin{equation}
		\boldsymbol{\Psi}_k
		=
		\boldsymbol{\Psi}_k^{\mathsf T},
		\qquad
		\boldsymbol{\Psi}_k^{\mathsf H}\boldsymbol{\Psi}_k
		=
		\mathbf I_{N_k+D_k}.
	\end{equation}
	Furthermore, its lower-left block satisfies
	\begin{align}
		[\boldsymbol{\Psi}_k]_
		{D_k+1:D_k+N_k,\,1:D_k}
		&=
		\overline{\mathbf Q}_k\boldsymbol{\Sigma}_k\overline{\mathbf P}_k^{\mathsf H}          \\
		&=
		2\widetilde{\mathbf U}_k.
	\end{align}
	Hence, the corresponding MiLAC combining matrix is
	\begin{equation}
		\mathbf U_k^{\mathrm m}
		=
		\frac{1}{2}
		[\boldsymbol{\Psi}_k]_
		{D_k+1:D_k+N_k,\,1:D_k}
		=
		\widetilde{\mathbf U}_k.
		\label{eq:milac_scaled_mmse}
	\end{equation}
	Therefore, the scaled and conjugated fully digital MMSE combiner can be exactly
	realized by a single-layer MiLAC.
	
	It remains to show that the scaling in
	\eqref{eq:scaled_mmse_receiver} does not affect the achievable
	rate. Define the effective desired-signal matrix and the
	post-combining interference-plus-noise covariance matrix of the
	fully digital MMSE combiner as
	\begin{align}
		\mathbf S_k^{\mathrm d}
		&=
		(\mathbf U_{k,\mathrm{MMSE}}^{\mathrm d})^{\mathsf H}
		\mathbf H_k\mathbf G_k,                                         \\
		\mathbf J_k^{\mathrm d}
		&=
		(\mathbf U_{k,\mathrm{MMSE}}^{\mathrm d})^{\mathsf H}
		\mathbf R_{-k}
		\mathbf U_{k,\mathrm{MMSE}}^{\mathrm d}.
	\end{align}
	Since
	$\mathbf U_k^{\mathrm m}
	=c_k\left(\mathbf U_{k,\mathrm{MMSE}}^{\mathrm d}\right)^*$, the corresponding
	MiLAC quantities satisfy
	\begin{align}
		\mathbf S_k^{\mathrm m}
		&=
		c_k\mathbf S_k^{\mathrm d},                                   \\
		\mathbf J_k^{\mathrm m}
		&=
		|c_k|^2\mathbf J_k^{\mathrm d}.
	\end{align}
	Because $c_k\neq0$, it follows that
	\begin{align}
		R_k^{\mathrm m}
		&=
		\log_2\det\left(
		\mathbf I_{D_k}
		+
		(\mathbf S_k^{\mathrm m})^{\mathsf H}
		(\mathbf J_k^{\mathrm m})^{-1}
		\mathbf S_k^{\mathrm m}
		\right)                                                        \\
		&=
		\log_2\det\left(
		\mathbf I_{D_k}
		+
		(\mathbf S_k^{\mathrm d})^{\mathsf H}
		(\mathbf J_k^{\mathrm d})^{-1}
		\mathbf S_k^{\mathrm d}
		\right)                                                        \\
		&=
		R_k^{\mathrm d}.
		\label{eq:receiver_rate_equivalence}
	\end{align}
	Thus, uniform nonzero scaling of the receive combiner scales the
	desired signal, interference, and post-combining noise by
	corresponding common factors, which cancel in the achievable-rate
	expression.
	
	Finally, the construction above maps a fully digital MMSE combiner to a feasible one-layer MiLAC combiner that attains exactly the same rate. Hence, the two receiver architectures achieve the same achievable rate for every user and, consequently, the same maximum sum-rate.
	\end{appendices}
	\bibliographystyle{IEEEtran}  
	\bibliography{reference}

\end{document}